\documentclass[12pt]{iopart}
\usepackage{graphicx}
\usepackage{bm}
\usepackage{setstack}
\usepackage{iopams}

\begin{document}

\title{Atomic four-wave mixing via condensate collisions}
\author{A. Perrin$^1$, C. M. Savage$^2$, D. Boiron$^1$, V. Krachmalnicoff$^1$, C. I. Westbrook$^1$ and K.V. Kheruntsyan$^3$}
\address{$^1$ Laboratoire Charles Fabry de l'Institut d'Optique, CNRS, Univ Paris-Sud,
Campus Polytechnique, RD128, 91127 Palaiseau Cedex, France }
\address{$^2$ ARC Centre of Excellence for Quantum-Atom Optics, Department of Physics,
Australian National University, Canberra ACT 0200, Australia}
\address{$^3$ ARC Centre of Excellence for Quantum-Atom Optics, School of Physical
Sciences, University of Queensland, Brisbane, QLD 4072, Australia}
\ead{\mailto{christoph.westbrook@institutoptique.fr},\mailto{karen.kheruntsyan@uq.edu.au}}

\begin{abstract}
We perform a theoretical analysis of atomic four-wave mixing via a collision
of two Bose-Einstein condensates of metastable helium atoms, and compare the
results to a recent experiment. We calculate atom-atom pair
correlations within the scattering halo produced spontaneously
during the collision. We also examine the expected relative number squeezing
of atoms on the sphere. The analysis includes first-principles quantum
simulations using the positive $P$-representation method. We develop a
unified description of the experimental and simulation results.
\end{abstract}

\submitto{\NJP}
\maketitle

\section{Introduction}

Recent years have seen the introduction of powerful new tools for studying
degenerate quantum gases. For example, on the experimental side correlation
measurements offer a new experimental probe of many-body effects
\cite{jeltes:07,Hellweg:03,esteve:06,rom:06,ottl:05,ritter:07,Hofferberth:07,MPC-Schellekens,MPC-Shimizu,Bloch,Dissociation-Greiner}. On the theoretical side, the challenges posed by the new experimental
techniques are being met by quantum dynamical simulations of large numbers
of interacting particles in realistic parameter regimes. These are becoming
possible due to the advances in computational power and improvements in
numerical algorithms (for recent examples, see~\cite%
{norrie:040401,Dissociation-Savage-etal-1,Deuar-Drummond-4WM}).

\begin{figure}[tbp]
\centering\includegraphics[height=5cm]{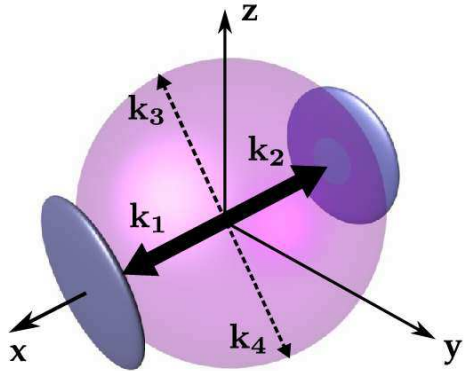} \caption{Schematic momentum space diagram of the atomic four-wave
mixing interaction. Optical Raman pulses generate untrapped
condensates with momenta $\mathbf{k}_1$ and $\mathbf{k}_2=-\mathbf{k}_1$
parallel
to the $x$-axis (dark disks). These undergo a four-wave mixing
interaction to produce correlated atomic pairs on a spherical shell
of radius $k_1$.} \label{fig1}
\end{figure}

In this paper we study metastable helium ($^{4}$He*),
which is currently
unique in \emph{quantum atom optics} in that it permits a comparison of
experimentally measured \cite{Orsay-collision-experiment} and theoretically calculated quantum correlations.
 This is one of the first examples in which
experimental measurements can be considered in the context of
first-principles calculations. Our goal in this paper is to confront a
theoretical analysis with the results of recent experiments on atomic four-wave mixing via a
collision of two Bose-Einstein condensates (BECs) of metastable $^{4}$He$%
^{\ast }$ atoms~\cite{Orsay-collision-experiment}. Figure~\ref{fig1} is a schematic
momentum space diagram of these experiments. Two condensates, whose atoms
have approximately equal but opposite momenta, $\mathbf{k}_1$ and $\mathbf{k}%
_2 \simeq-\mathbf{k}_1$, interact by four-wave mixing, while they
spatially overlap, to produce correlated atomic pairs with approximately
equal but opposite momenta, $\mathbf{k}_3$ and $\mathbf{k}_4$, satisfying
momentum conservation, $\mathbf{k}_1 + \mathbf{k}_2 = \mathbf{k}_3 + \mathbf{%
k}_4$. Figure~\ref{fig1} corresponds to the experimental data shown in figure~2 of \cite%
{Orsay-collision-experiment}, since after time-of-flight expansion, atomic
momentum is mapped into atomic position.

We perform first-principles quantum simulations of the collision dynamics
using the positive $P$-representation method \cite%
{Drummond-Gardiner,Steel-Olsen,Drummond-Corney}. The advantage of
this method is that given the Hamiltonian of the interacting
many-body system, no additional approximations are imposed to
simulate the quantum dynamics governed by the Hamiltonian. The
drawback on the other hand, is that it usually suffers from large
sampling errors and the boundary term problem \cite{Gilchrist} as
the simulation time $t_{\mathrm{sim}}$ increases, eventually leading
to diverging results.

An empirically estimated upper bound for the positive-$P$ simulation time
(with controllable sampling error) of condensates with $s$-wave scattering
interactions is given approximately by \cite{Deuar-Drummond}%
\begin{equation}
t_{\mathrm{sim}}\lesssim 2.5m(\Delta V)^{1/3}/[4\pi \hbar a \rho_{0}^{2/3}],
\end{equation}%
where $m$ is the atom mass, $a$ is the $s$-wave scattering length, $\rho
_{0} $ is the condensate peak density, and $\Delta V=\Delta x\Delta y\Delta
z $ is the volume of the elementary cell of the computational lattice, with
lattice spacings of $\Delta x$, $\Delta y$, and $\Delta z$. Applying this formula to
metastable helium, we see that this is a particularly challenging case among
commonly condensed atoms due to its small atomic mass and relatively large
scattering length. Our simulations are restricted to short
interaction times (typically $\lesssim 25$ $\mu $s), which are about $6$ times
shorter than the experimental interaction  time of~\cite%
{Orsay-collision-experiment}. Despite this, our positive-$P$ simulations
provide useful insights into the experimental observations; in addition,
they can serve as benchmarks for approximate theoretical methods (such as
the Hartree-Fock-Bogoliubov method \cite{zin:05,zin:06,morgan,yurovsky:02}) to establish
the range of their validity.

We calculate atom pair correlations within the scattering halo produced
spontaneously during the collision (see figure~\ref{fig1}). The scattering
halo is a spherical shell in momentum space. In the limit of small
occupation of the scattered modes, the $s$-wave nature of the collisions
ensures an approximately uniform atom population over the halo. We consider
the strength and the width of the correlation signal, as well as the
momentum width of the halo. We also analyze relative atom number squeezing
and the violation of the classical Cauchy-Schwartz inequality.

In Sec.~2 of this paper we will summarize the experimental results we wish
to analyze. In Sec.~3 we discuss order of magnitude estimates.
In Sec.~4 we describe simulations using the positive $P$-representation
method, and in Sec.~5 we discuss the results of our simulations. Sec.~6
summarizes our work.

\section{Summary of experimental results}

\subsection{Overview of the experiment}

The starting point of the experiment is a $^{4}$He* condensate of $10^{4}$
to $10^{5}$ atoms confined in a magnetic trap whose frequencies are: $\omega
_{x}/2\pi =47$ Hz and $\omega _{y}/2\pi =\omega _{z}/2\pi =1150$ Hz. A
sudden Raman outcoupling drives the trapped $^{4}$He* from the $m_{x}=1$
Zeeman sublevel into the magnetic field insensitive state $m_{x}=0$. 
~\cite{Orsay-collision-experiment}. The Raman transition also splits the
initial ($m_{x}=1$) condensate into two roughly equally populated
condensates with opposite velocities along the $x$ direction.
The magnitude of each velocity is equal to the recoil velocity $v_{r}=9.2$
cm/s, defined by the momentum of the photons used to create the colliding
condensates $\hbar k_{r}$, $k_{r}=5.8\times 10^{6}$ m$^{-1}$. The relative
velocity $2v_{r}$ of the two condensates is about $8$ times higher than the
speed of sound $c_{s}=\sqrt{\mu /m}$ of the initial condensate, ensuring
that elementary excitations of the condensates correspond to free particles.

During the separation of the condensates, elastic collisions
occurring between atoms with opposite velocities scatter a small
fraction ($5\%$) of the total initial atom number into the halo.
The
system is shown in three-dimensions in an accompanying video of the
experimental results after a $320$~ms time of flight~\footnote{A 3 dimensional, animated rendition of the atomic positions 320 ms after
release from the trap.
The vertical positions are derived from the arrival times as described in
\cite{Orsay-collision-experiment}.
Each point corresponds to the detection of one atom and the animation shows
the sum of 50 separate runs.
The ellipsoids at the sides are the colliding condensates.
The ellipsoids at the top and bottom result from imperfect Raman
polarizations and stimulated atomic 4 wave mixing (see \cite{Orsay-collision-experiment}).
The 4 condensates are excluded from the analysis given in the text.}.
For the purposes of this paper, the experiment consists in the
acquisition of the three dimensional positions of the particles
scattered into the collision halo after the time of flight. This
information permits the reconstruction of the 3D momentum vectors of
the individual particles after they have ceased interacting with
each other.

\subsection{Main results}

Knowledge of the momentum vectors in turn permits the construction of
two-particle correlation functions in momentum space. The correlation function
shows features  for
particles traveling both back to back and collinearly. The back-to-back
correlation results from binary, elastic collisions between atoms,
whereas the collinear correlation is a two particle interference
effect involving members of two different pairs: a Hanbury Brown-Twiss
correlation~\cite{Moelmer-Orsay}. Both correlation functions are anisotropic
because of the anisotropy of the initial colliding condensates.

To quantify these correlations, we first introduce the unnormalized
normally-ordered second-order correlation function between the densities at
two points in momentum space,
\begin{equation}
G^{(2)}(\mathbf{k}_{1},\mathbf{k}_{2})=\langle :\hat{n}(\mathbf{k}_{1})\hat{n%
}(\mathbf{k}_{2}):\rangle .
\end{equation}%
Here, $\hat{n}(\mathbf{k})=\hat{a}^{\dagger }(\mathbf{k})\hat{a}(\mathbf{k})$
is the momentum density operator, $\hat{a}^{\dagger }(\mathbf{k})$ are $\hat{%
a}(\mathbf{k})$ are the Fourier transforms of the field creation and
annihilation operators $\hat{\Psi}^{\dagger }(\mathbf{x})$ and $\hat{\Psi}(%
\mathbf{x})$, and the colons :: stand for normal ordering of the operators
according to which all creation operators stand to the left of the
annihilation operators, so that
\begin{equation}
\langle :\hat{n}(\mathbf{k}_{1})\hat{n}(\mathbf{k}_{2}):\rangle =\langle
\hat{a}^{\dagger }(\mathbf{k}_{1})\hat{a}^{\dagger }(\mathbf{k}_{2})\hat{a}(%
\mathbf{k}_{2})\hat{a}(\mathbf{k}_{1})\rangle .
\end{equation}

Because of a low data rate, the correlation measurements must be
averaged over the entire collision sphere to get statistically
significant results. The average \emph{collinear} (CL) second-order
correlation as a function of the relative displacement $\Delta
k_{i}$ in the $k_{i}$-direction ($i=x,y,z$) is defined as
\begin{equation}
g_{CL}^{(2)}(\Delta k_{i})=\frac{\int\limits_{\mathcal{D}}d^{3}\mathbf{k}~G^{(2)}(\mathbf{k},\mathbf{k+e}_{i}\Delta k_{i})}{\int\limits_{\mathcal{D}%
}d^{3}\mathbf{k}~\langle \hat{n}(\mathbf{k})\rangle \langle \hat{n}(\mathbf{%
k+e}_{i}\Delta k_{i})\rangle },  \label{g2-CL-av}
\end{equation}%
where $\mathbf{e}_{i}$ is the unit vector in the $k_{i}$ direction.
The normalization of $g_{CL}^{(2)}(\Delta k_{i})$ ensures that for
uncorrelated densities $g_{CL}^{(2)}(\Delta k_{i})=1$. The
integration domain $\mathcal{D}$ in~(\ref{g2-CL-av}) selects a
certain region of interest in momentum space and can be defined, for
example, to contain only a narrow spherical shell and to eliminate
the initial colliding condensates. Due to the averaging, the
dependence of the correlation functions on the direction
$\mathbf{k}$ is lost.

The average \emph{back-to-back} (BB) correlation function
$g_{BB}^{(2)}(\Delta k_{i})$ between two diametrically opposite
points, one of which is additionally displaced by $\Delta k_{i}$ in
the $k_{i}$-direction, is defined similarly to $g_{CL}^{(2)}(\Delta
k_{i})$:
\begin{equation}
g_{BB}^{(2)}(\Delta k_{i})=\frac{\int\limits_{\mathcal{D}}d^{3}\mathbf{k}~G^{(2)}(\mathbf{k},-\mathbf{k+e}_{i}\Delta k_{i})}{\int\limits_{\mathcal{D}%
}d^{3}\mathbf{k}~\langle \hat{n}(\mathbf{k})\rangle \langle \hat{n}(-\mathbf{%
k+e}_{i}\Delta k_{i})\rangle }.  \label{g2-BB-av}
\end{equation}

The experimental observations can be summarized as follows. The width of
both correlation functions along the axial direction of the condensate, the $%
x$-axis, is limited by the resolution of the detector and hence contains
little information about the collision. In the radial direction (with
respect to the symmetry of the colliding condensates), one observes a peak
which can be fitted to a Gaussian function with rms widths $\sigma _{y,z}^{CL}$
and $\sigma _{y,z}^{BB}$ for collinear and back-to-back cases respectively.
The experimental results are summarized in the following table
\begin{equation}
\begin{tabular}{|c|c|c|}
\hline
$\sigma _{y,z}^{BB} /k_r$ & $\sigma _{y,z}^{CL} /k_r$ & $\sigma
_{y,z}^{CL}/\sigma _{y,z}^{BB}$ \\ \hline
$0.081\pm 0.004$ & $0.069\pm 0.008$ & $0.85\pm 0.15$ \\ \hline
\end{tabular}
\label{g2-expt}
\end{equation}

\begin{figure}[tbp]
\centering\includegraphics[height=5cm]{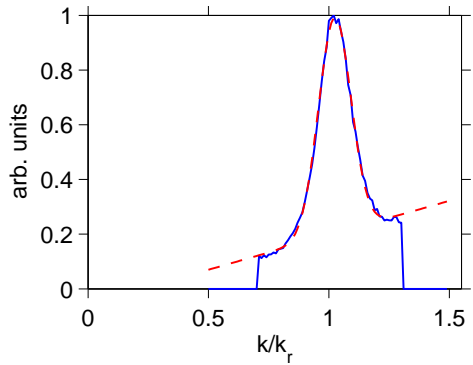} \caption{Cross section of the scattering halo.
A sloped background is present due to thermal atoms in the trap.
This background has been fit to a straight line and subtracted in order to
estimate the rms width, $\delta k\simeq 0.067k_{r}$.} \label{figradius}
\end{figure}

One can also use the data to deduce the averaged radial width
$\delta k$ of the scattering halo. Figure~\ref{figradius} shows a
cross section of the halo, averaged over all accessible scattering
angles. The presence of the unscattered condensates prevents
observation of the shell along the $x$-axis, but along the
accessible directions we find $\delta k\simeq 0.067k_{r}$.

\section{Qualitative analysis}\label{qualitative-analysis}

In this section we discuss some simple, analytical estimates of the measured
quantities. In later sections we will do more precise, numerical
calculations which will verify the results of this section.

\subsection{Width of the pair correlation functions}

As discussed in~\cite{Orsay-collision-experiment}, the width of the
back-to-back and collinear correlation functions should be on the order of
the momentum width of the initial condensate, which in turn is proportional
to the inverse width of its spatial profile. For a Gaussian density profile
of the initial condensate in position space $\rho (\mathbf{x)}=\langle \hat{%
\Psi}^{\dagger }(\mathbf{x})\hat{\Psi}(\mathbf{x})\rangle =\rho
_{0}\exp [-\sum_{i=x,y,z}r_{i}^{2}/(2w_{i}^{2})]$, and therefore a
Gaussian
density distribution in momentum space, $n(\mathbf{k)}=\langle \hat{n}(%
\mathbf{k})\rangle \propto \exp [-\sum_{i=x,y,z}k_{i}^{2}/(2\sigma
_{i}^{2})] $, with $\sigma _{i}=1/w_{i}$, an approximate theoretical
treatment based on a simple ansatz for the pair wavefunction
predicts a Gaussian dependence of the back-to-back (BB) and
collinear (CL) correlation
functions on the relative wavevectors $\Delta k_{i}$ \cite{Moelmer-Orsay}:%
\begin{eqnarray}
G^{(2)}(\mathbf{k},-\mathbf{k+n}_{i}\Delta k_{i}) &\propto &\exp \left( -%
\frac{\Delta k_{i}^{2}}{2(\sigma _{i}^{BB})^{2}}\right) , \\
G^{(2)}(\mathbf{k},\mathbf{k+n}_{i}\Delta k_{i}) &\propto &\exp \left( -%
\frac{\Delta k_{i}^{2}}{2(\sigma _{i}^{CL})^{2}}\right) .
\end{eqnarray}

The widths of the back-to-back ($\sigma _{i}^{BB}$) and collinear
($\sigma _{i}^{CL}$) correlations are related to the momentum-space
width $\sigma _{i} $ of the initial (source) condensate via
\cite{Moelmer-Orsay}
\begin{eqnarray}
\sigma _{i}^{BB}/\sigma _{i} &=&\sqrt{2},  \label{corr_widths_gauss_CL} \\
\sigma _{i}^{CL}/\sigma _{i} &=&2,  \label{corr_widths_gauss_BB}
\label{ansatz-result}
\end{eqnarray}%
and therefore the width of the back-to-back correlation is $\sqrt{2}$ times
smaller than the width of the collinear correlation. 

In Sec.~\ref{sect:main-example} the initial momentum-space widths are found
to be $\sigma _{x}=0.0025k_{r}$ and $\sigma _{y,z}=0.055k_{r}$, assuming $%
N=9.84\times 10^{4}$ atoms. Expressing the experimentally measured
widths in units of $\sigma _{y,z}$, we can rewrite~(\ref{g2-expt}) as
\begin{equation}
\begin{tabular}{|c|c|c|}
\hline $\sigma_{y,z}^{BB} /\sigma_{y,z}$ & $\sigma_{y,z}^{CL}
/\sigma_{y,z}$ & $\sigma _{y,z}^{CL}/\sigma_{y,z}^{BB}$ \\ \hline
$1.47\pm 0.07$ & $1.25\pm 0.15$ & $0.85\pm 0.15$ \\ \hline
\end{tabular}
\label{g2-expt-2}
\end{equation}
and therefore,~(\ref{corr_widths_gauss_CL}) is in agreement with the
measured width of the radial back-to-back correlation function,
whereas~(\ref{corr_widths_gauss_BB}) overestimates the
width of the collinear correlation function by almost $60\%$. As
we show below, first-principles simulations using the positive-$P$
method and incorporating atom-atom interactions result in widths
which are closer to the experimental values.

The discrepancy between the two theoretical approaches (which apparently is
larger for the collinear correlations than for the back-to-back ones) comes
mostly from the
fact that the above calculation is made for a Gaussian shape of the initial
BEC density profile, whereas in practice and in the positive-$P$ simulations
the spatial density of a harmonically trapped condensate is closer to an
inverted parabola (as in the Thomas-Fermi limit) rather than to a Gaussian.
An alternative theoretical model \cite{MagnusOgren}, based on the undepleted
source condensate approximation and a numerical solution to the linear
operator equations of motion for scattered atoms, also confirms that for
short times the momentum-space correlation widths are narrower if the source
condensate has a parabolic spatial density profile, compared to the case of
a Gaussian density profile.

\subsection{Width of the scattered halo}

\label{width_scatt_sphere}

A second, experimentally accessible quantity in a BEC collision is
the width $\hbar \delta k_{i}$ in momentum space of the halo on
which the
scattered atoms are found. Clearly the momentum spread $\sigma _{i}$ (in $%
i=x $, $y$ or $z$ direction) of the colliding condensates imposes a
minimum width
\begin{equation}
\delta k_{i}\gtrsim \sigma _{i}.  \label{sigma}
\end{equation}%
This limit suggests that the halo could be anisotropic. As noted
above however, the experiment in~\cite{Orsay-collision-experiment} is
not highly sensitive to such an anisotropy, and measures the width chiefly
in the $y,z$-directions.

Other physical considerations also affect this width, and suggest that the
halo should rather be isotropic, in which case we can drop the
index from $\delta k$. Here we discuss two mechanisms that impose a finite
radial width on the halo.

If the pairs are produced during a finite time interval $\Delta t$,
the total energy of the pair is necessarily broadened by $\hbar
/\Delta t$. This is true even if the relative momentum is well
defined. For a mean $k$-vector $k_{r}$, the finite interaction time
between the colliding BECs results in a broadening of
\begin{equation}
\delta k\simeq {\frac{m}{{\hbar k_{r}\Delta t}},}  \label{delta-k}
\end{equation}
where we assumed $\delta k/k_{r}\ll 1$. In the experiment, the collision
time is sufficiently long that the above effect does not impose a limitation
on the width of the sphere. In the positive-$P$ simulations however,
numerical stability problems limit the maximum collision time that can be
simulated, as discussed in Sec. \ref{Numerical results}, and this time does
indeed impose a width on the halo. For short collision times,
where the scattering is in the spontaneous regime, our numerical results for
the width $\delta k$ are in good agreement with the simple estimate of~\Eref{delta-k}.

For long collision times it can happen that so many atoms are
scattered that Bose enhancement and stimulated effects become
important. In this case the width of the scattering shell can be
estimated by a slightly more involved approximate approach based on
analytic solutions for the uniform system within the undepleted
\textquotedblleft pump\textquotedblright\ (source condensate)
approximation \cite{Bach-Trippenbach-Rzazewski-2002}. Under this
approximation, the present system is equivalent to the dissociation
of a condensate of
molecular dimers studied in detail in~\cite%
{Dissociation-Savage-etal-1,Dissociation-Twin-Beams,Dissociation-Davis-etal}.
The latter system in turn is analogous to parametric down-conversion
in optics \cite{Walls-Milburn}. The details of the approximate
solutions, common to condensate collisions and molecular
dissociation, and the relationship between them are given
in~\ref{sect:undepletedpump}. The resulting width of the halo found
from this approach is
\begin{equation}
\delta k\simeq \frac{4\pi a \rho _{0}}{k_{r}}.  \label{width-stim}
\end{equation}%
We see that in this regime, the width is proportional to the scattering
length $a$ and the peak density $\rho _{0}$, but it no longer depends on the
collision duration.

The physical interpretation of~\Eref{width-stim} is that with the
stronger effective coupling (or nonlinearity) $a \rho_{0}$, one can
excite and amplify spectral components that are further detuned from
the exact resonance condition $\hbar \Delta_{k}=0$ (or further
``phase mismatched''). The inverse dependence on collision momentum
$k_{r}$ can be understood via the quadratic dependence of the energy
on momentum $k$: to get the same excitation at a given energy offset
$\hbar \Delta_{k}$, (\ref{eq:resonance}), one requires smaller
absolute momentum offset $\delta k$ at larger $k_{r}$ than at small
$k_{r}$.

Positive-$P$ simulations covering the transition from the spontaneous
to stimulated regimes are available for $^{23}$%
Na condensate collisions as in \cite{Deuar-Drummond-4WM}. The
numerical results in this case are in agreement with the simple
analytic estimate of \Eref{width-stim}. More specifically, we find
that for collision durations between $300$ $\mu$s and $640$ $\mu$s the
actual numerical results for the width of the spherical halo vary,
respectively, between $\delta k/k_{r}\simeq 0.13$ and $\delta
k/k_{r}\simeq 0.087$, whereas \Eref{width-stim} predicts
$\delta k/k_{r}\simeq 0.096$.

For $^{4}$He$^{\ast }$, on the other hand, the small mass and the
larger scattering length of $^{4}$He$^{\ast }$ atoms limit the
maximum simulation time to $t_{\mathrm{sim}}\lesssim $ $25$ $\mu $s.
This is far from the stimulated regime and therefore we do not have
a direct comparison of the numerical results with \Eref{width-stim}.
The experiment is also not in the stimulated regime. We are
nevertheless tempted by the numerical $^{23}$Na result
 to extrapolate \Eref%
{width-stim} to $^{4}$He$^{\ast }$ BEC collisions in the long time
limit and we obtain $\delta k/k_{r}\simeq 0.05$. Adding this width
in quadrature to the momentum width of the initial condensate,
$\sigma_{y,z} \simeq0.055 k_r$, gives $\sqrt{ (
0.05 k_r )^2 +(0.055 k_r)^2} = 0.074k_r$,
not far from the experimentally observed radial momentum width of $\delta k \simeq
0.067k_{r}$.
We thus suggest that the mechanism leading to \Eref{width-stim} may play
a role in the experiment.

\section{Model}

The effective field theory Hamiltonian governing the dynamics of the
collision of BECs is given by%
\begin{equation}
\hat{H}=\int d\mathbf{x}\left\{ \frac{\hbar ^{2}}{2m}|\mathbf{\nabla }\hat{%
\Psi}|^{2}+\frac{\hbar U_{0}}{2}\hat{\Psi}^{\dag }\hat{\Psi}^{\dag }\hat{\Psi%
}\hat{\Psi}\right\} ,
\end{equation}%
where $\hat{\Psi}(\mathbf{x},t)$ is the field operator with the usual
commutation relation $[\hat{\Psi}(\mathbf{x},t),\hat{\Psi}^{\dagger }(%
\mathbf{x}^{\prime },t)]=\delta ^{(3)}(\mathbf{x-x}^{\prime })$, $m$ is the
atomic mass, the first term is the kinetic energy, and the second term
describes the $s$--wave scattering interactions between the atoms. The
trapping potential for preparing the initial condensate before the collision
is omitted since we are only modeling the dynamics of the outcoupled
condensates in free space. The use of the effective delta function
interaction potential $U(x-y)=U_{0}\delta (x-y)$ assumes a UV momentum
cutoff $k^{\max }$. In our numerical simulations the momentum cutoff is
imposed explicitly via the finite computational lattice. If the lattice
spacings\ ($\Delta x$, $\Delta y$, $\Delta z$) in each spatial dimensions
are chosen to be much larger than the $s$--wave scattering length $a$, then
the respective momentum cutoffs satisfy $k_{x,y,z}^{\max }\ll 1/a$. In this
case the coupling constant $U_{0}$ is given by the familiar expression $%
U_{0}\simeq 4\pi \hbar a/m$ \cite{Abrikosov} without the need for explicit
renormalization.

To model the dynamics of quantum fields describing the collision of two
BECs, we use the positive $P$--representation approach \cite%
{Drummond-Gardiner}. In this approach the quantum field operators $\hat{\Psi}%
(\mathbf{x},t)$ and $\hat{\Psi}^{\dagger }(\mathbf{x},t)$ are represented
by\ two\ complex stochastic $c$--number fields $\Psi (\mathbf{x},t)$ and $%
\tilde{\Psi}(\mathbf{x},t)$ whose dynamics is governed by the following
stochastic differential equations \cite{Deuar-Drummond-4WM}:%
\numparts
\begin{eqnarray}
\frac{\partial \Psi (\mathbf{x},t)}{\partial t} &=&\frac{i\hbar }{2m}\mathbf{%
\nabla }^{2}\Psi -iU_{0}\tilde{\Psi}\Psi \Psi +\sqrt{-iU_{0}\Psi ^{2}}%
\;\zeta _{1}(\mathbf{x},t),  \\
\frac{\partial \tilde{\Psi}(\mathbf{x},t)}{\partial t} &=&-\frac{i\hbar }{2m}%
\mathbf{\nabla }^{2}\tilde{\Psi}+iU_{0}\Psi \tilde{\Psi}\tilde{\Psi}+\sqrt{%
iU_{0}\tilde{\Psi}^{2}}\;\zeta _{2}(\mathbf{x},t).  
\end{eqnarray}%
\endnumparts
Here, $\zeta _{1}(\mathbf{x},t)$ and $\zeta _{2}(\mathbf{x},t)$ are real
independent noise sources with zero mean, $\langle \zeta _{j}(\mathbf{x}%
,t)\rangle =0$, and the following nonzero correlation:
\begin{equation}
\langle \zeta _{j}(\mathbf{x},t)\zeta _{k}(\mathbf{x}^{\prime },t^{\prime
})\rangle =\delta _{jk}\delta ^{(3)}(\mathbf{x}-\mathbf{x}^{\prime })\delta
(t-t^{\prime }).
\end{equation}%
The stochastic fields $\Psi (\mathbf{x},t)$ and $\tilde{\Psi}(\mathbf{x},t)$
are independent of each other [$\tilde{\Psi}(\mathbf{x},t)\neq \Psi ^{\ast }(%
\mathbf{x},t)$] except in the mean, $\langle \tilde{\Psi}(\mathbf{x}%
,t)\rangle =\langle \Psi ^{\ast }(\mathbf{x},t)\rangle $, where the brackets
$\langle \ldots \rangle $ refer to stochastic averages with respect to the
positive $P$--distribution function. In numerical realizations, this is
represented by an ensemble average over a large number of stochastic
realizations (trajectories). Observables described by quantum mechanical
ensemble averages over normally-ordered operator products have an exact
correspondence with stochastic averages over the fields $\Psi (\mathbf{x},t)$
and $\tilde{\Psi}(\mathbf{x},t)$:%
\begin{equation}
\langle \lbrack \hat{\Psi}^{\dagger }(\mathbf{x},t)]^{m}[\hat{\Psi}(\mathbf{x%
}^{\prime },t)]^{n}\rangle =\langle \lbrack \tilde{\Psi}(\mathbf{x}%
,t)]^{m}[\Psi (\mathbf{x}^{\prime },t)]^{n}\rangle .
\end{equation}

The initial condition for our simulations is a coherent state of a trapped
condensate, modulated with a standing wave that imparts initial momenta $\pm
k_{r}$ (where $k_{r}=mv_{r}/\hbar $ and $v_{r}$ is the collision velocity)
in the $x$ direction,%
\begin{equation}
\Psi (\mathbf{x},0)=\langle \hat{\Psi}(\mathbf{x},0)\rangle =\sqrt{\rho _{0}(%
\mathbf{x})/2}\left( e^{ik_{r}x}+e^{-ik_{r}x}\right) ,
\end{equation}%
with $\tilde{\Psi}(\mathbf{x},0)=\Psi ^{\ast }(\mathbf{x},0)$. Here, $\rho
_{0}(\mathbf{x})$ is the density profile given by the ground state solution
to the Gross-Pitaevskii equation in imaginary time. The above initial
condition models a sudden Raman outcoupling of a BEC of trapped $^{4}$He$%
^{\ast }$ atoms in the $m_{x}=1$ sublevel into the magnetic field
insensitive state $m_{x}=0$, using two horizontally counter-propagating
lasers and a third vertical laser \cite{Orsay-collision-experiment}. In this
geometry, the Raman transitions split the initial ($m_{x}=1$) condensate
into two equally populated condensates and simultaneously impart velocities
of $\pm v_{r}$ onto the two halves. As a result the two outcoupled
condensates undergo a collision and expand in free space. Accordingly, in
our dynamical simulations, the field $\hat{\Psi}(\mathbf{x},t)$ represents
the atoms in the untrapped state $m_{x}=0$, having the $s$--wave scattering
length of $a_{00}=5.3$ nm (\cite{Orsay-collision-experiment} and references therein), while the
initial density profile $\rho _{0}(\mathbf{x})$ refers to that of the
trapped atoms in the $m_{x}=1$ state having the scattering length of $%
a_{11}=7.51$ nm \cite{moal:06}. The same distinction in terms of the
scattering length in question applies to the definition of the interaction
strength $U_{0}\simeq 4\pi \hbar a/m$, in which $a$ has to be understood as $%
a_{11}$ for the trapped condensate or as $a_{00}$ for the outcoupled cloud.

In our simulations we assume for simplicity that the outcoupling
from the trapped $m_{x}=1$ state is $100\%$ efficient, in which case
the entire population is transferred into the $m_{x}=0$ state and
therefore we have to only model $s$-wave scattering interactions
between the atoms in the $m_{x}=0$ state. In the experiment, on the other
hand, the transfer efficiency is only about $60\%$ and therefore the
collisions between the atoms in the $m_{x}=0$ and $m_{x}=1$ are not
completely negligible and maybe responsible for some of the
deviations between the present theoretical results and the experimental
observations.

\section{Results and discussion}

\label{Numerical results}

\subsection{Main numerical example}

\label{sect:main-example}

Here we present the results of positive-$P$ numerical simulations of
collisions of two condensates of $^{4}$He$^{\ast }$ atoms ($m\simeq
6.65\times 10^{-27}$ kg) as in the experiment of~\cite%
{Orsay-collision-experiment}. The key parameters in our main
numerical example are the collision velocity, $v_{r}=9.2$ cm/s,
 and the peak density
of the initial trapped condensate, $\rho _{0}=2.5\times 10^{19}$
m$^{-3}$.
The trap frequencies are matched exactly with the experimental values, $%
\omega _{x}/2\pi =47$ Hz and $\omega _{y}/2\pi =\omega _{z}/2\pi =1150$ Hz.
The $s$-wave scattering length for the magnetically trapped atoms in the $%
m_{x}=1$ sublevel is $a_{11}=7.5$ nm; the $s$-wave scattering length for the
outcoupled atoms in the $m_{x}=0$ sublevel is $a_{00}=5.3$~nm. Other 
simulation parameters are given in Appendix \ref{sect:parameters}.

The initial state of the trapped condensate is found via the solution of the
Gross-Pitaevskii equation in imaginary time. Given the above trap
frequencies and the peak density as a target, we find that the total number
of atoms in the main example is $N=9.84\times 10^{4}$. With these
parameters, the average kinetic energy of colliding atoms is $%
E_{kin}/k_{B}=mv_{r}^{2}/2k_{B}\simeq 2.0\times 10^{-6}$ K, which is about $%
7.4$ times larger than the mean-field energy of the initial condensate $%
E_{MF}/k_{B}=4\pi \hbar ^{2}a_{11}\rho _{0}/mk_{B}\simeq 2.7\times 10^{-7}$
K.

\begin{figure}[tbp]
\centering\includegraphics[height=5cm]{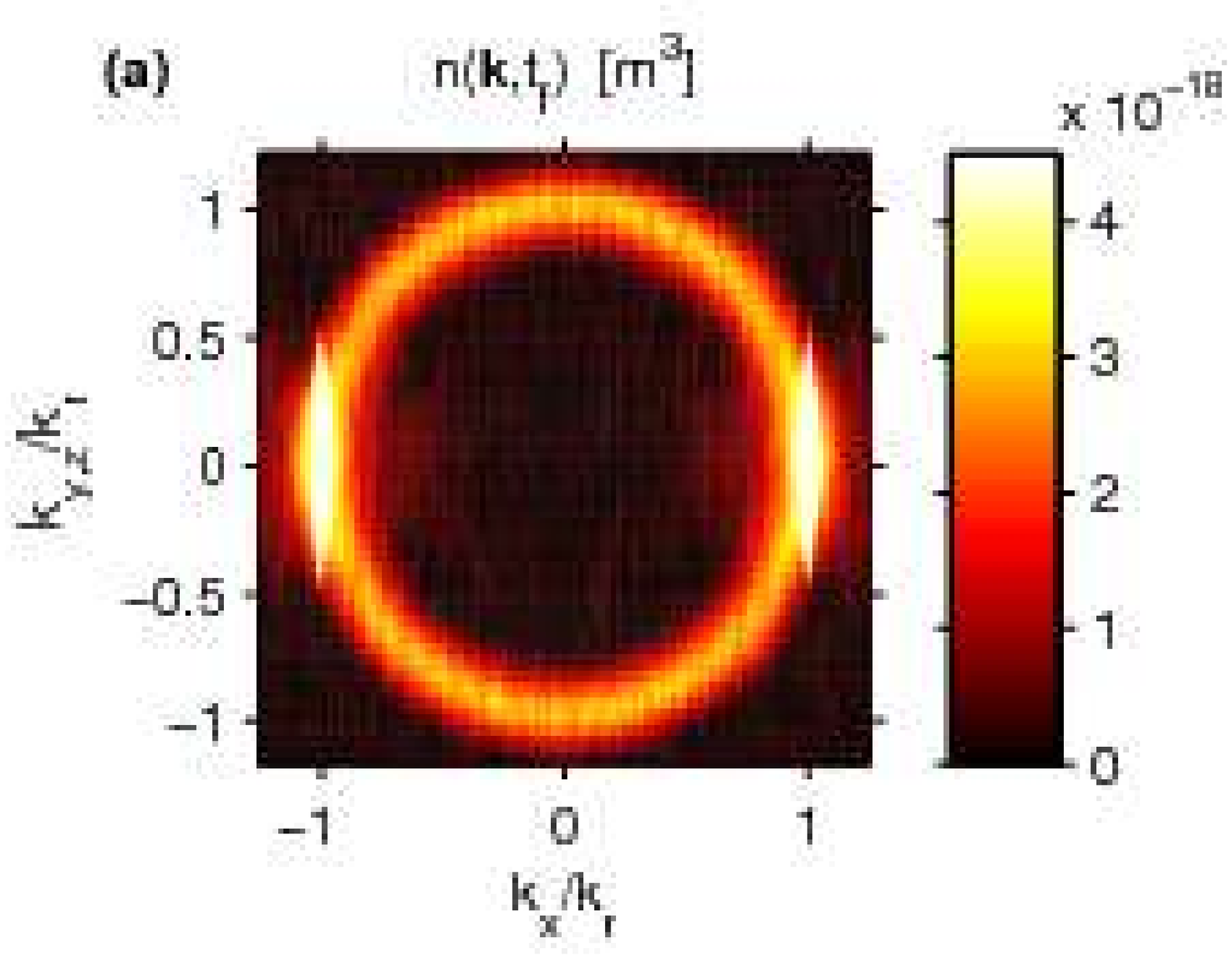}
\includegraphics[height=5cm]{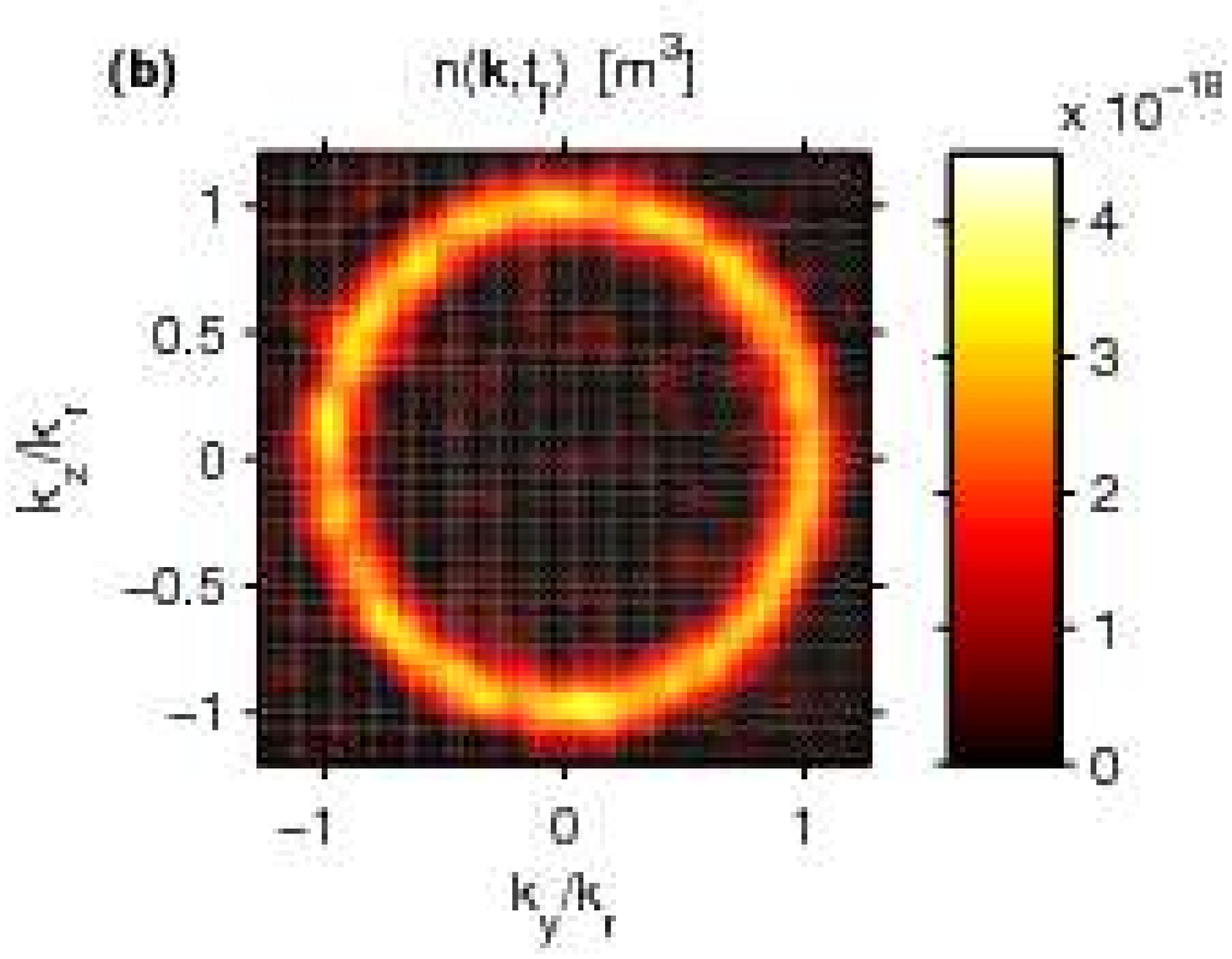}
\caption{Slices through $k_{z}=0$ (a) and $k_{x}=0$ (b) of the 3D atomic
density distribution in momentum space $n({\mathbf{k}},t_{f})$ after $%
t_{f}=25$ $\protect\mu $s collision time. Due to the symmetry in the
transverse direction (orthogonal to $x$), the average density through $%
k_{y}=0$ coincides with that of $k_{z}=0$. The color scale is chosen to
clearly show the halo of spontaneously scattered atoms and
cuts off the high-density peaks of the two colliding condensates (shown in
white on the left panel).}
\label{fig2}
\end{figure}

The duration of simulation in the main example is $t_{f}=25$ $\mu $s. This
is considerably  smaller than the estimated duration of collision in the
experiment,  $140$ $\mu $s (see \ref%
{sect:collisionduration}). The number of scattered atoms in our numerically
simulated example at $t_{f}=25$ $\mu $s is $\sim 1750$, representing $\sim
1.8$\% of the total number of atoms in the initial BEC. Operationally, the
fraction of scattered atoms is determined as the total number of atoms
contained within the scattering halo (see figure~\ref{fig2} showing
two orthogonal slices through the momentum density distribution) after
eliminating the regions of momentum space occupied by the two colliding
condensates. We implement the elimination by simply discarding the data
points corresponding to $|k_{x}|>0.99k_{r}$, which fully contain the
colliding condensates. This cuts off a small  fraction of the
scattered atoms as well, but the procedure is simple to implement
operationally and is unambiguous.

\begin{figure}[tbp]
\centering\includegraphics[height=5cm]{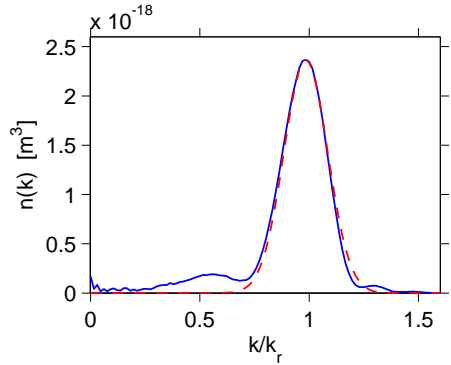}
\caption{Angle averaged (radial) momentum distribution $n(k)$ of the
 scattered atoms (solid line) and a simple Gaussian fit (dashed line)
used to define the radial width $\delta k=0.10k_{r}$ of the halo
around the peak momentum $k_{0}=0.98k_{r}$ (see text).}
\label{fig3}
\end{figure}

In order to compare our calculated fraction of scattered atoms at
$t_{f}=25$ $\mu $s with the experimentally measured fraction of
$5$\% at the end of collision at $\sim 140$ $\mu $s, we first note
that these time scales are relatively short and correspond to the
regime of spontaneous scattering. The number of scattered atoms
increases approximately linearly with time, therefore our calculated
fraction of $1.8$\% can be extrapolated to about $10$\% to
correspond to the expected fraction at $\sim 140$ $\mu $s. Next, one
has to scale this value by a factor $0.6^2$ to account for the fact
that in the experiment only $60\%$ of the initial atom number was
transferred to the $m_{x}=0$ state of the colliding condensates.
Accordingly, our theoretical estimate of $10$\% should be
proportionally scaled down to $4\%$ conversion, in good agreement
with the experimentally estimated fraction of $5$\% (see
also~\ref{sect:collisionduration}).

In figure~\ref{fig3} we plot the radial momentum distribution of
scattered atoms (solid line), obtained after angle averaging of the
full 3D distribution within the region $|k_{x}|\leq 0.8k_{r}$. The
numerical result is fitted with a Gaussian $\propto \exp
[-(k-k_{0})^{2}/(2\delta k^{2})]$
(dashed line), centered at $k_{0}=0.98k_{r}$ and having the radial width of $%
\delta k=0.10k_{r}\simeq 5.8\times 10^{5}$ m$^{-1}$, where
$k=|\mathbf{k}|$. The fitted radial width of $\delta k=0.10k_{r}$ of
the numerical simulation is in reasonable agreement with the simple
estimate of \Eref{delta-k}, which gives $\delta k\simeq
0.075k_{r} $ for $\Delta t=25$ $\mu$s.

\begin{figure}[tbp]
\centering\includegraphics[height=5cm]{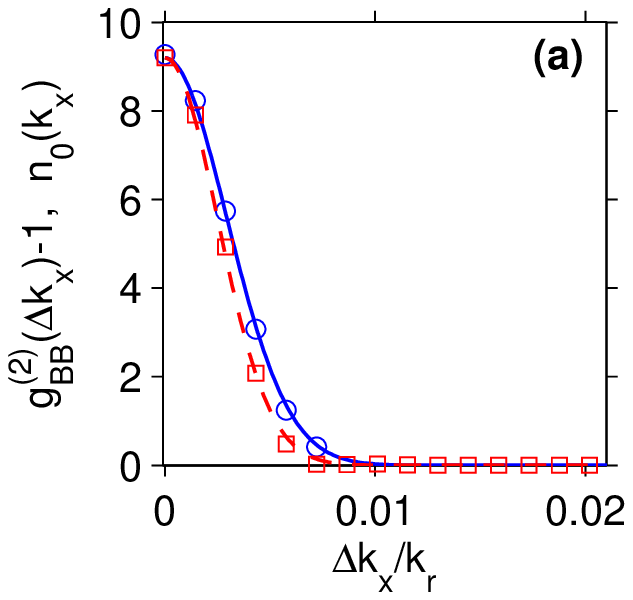}%
\includegraphics[height=5cm]{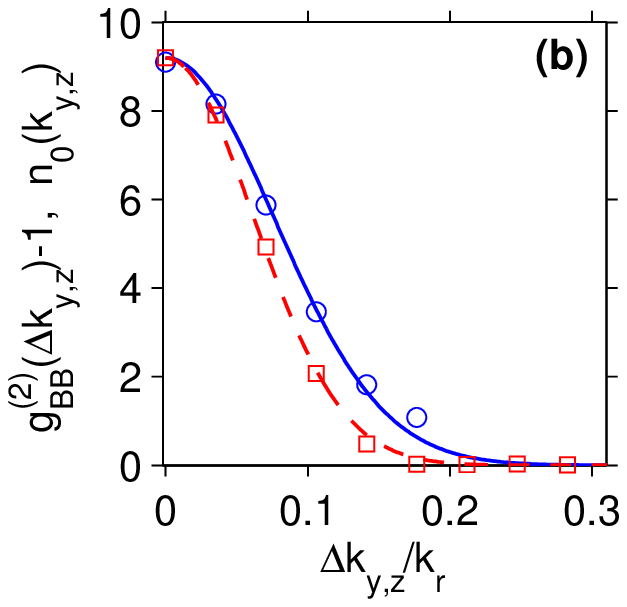} %
\includegraphics[height=5cm]{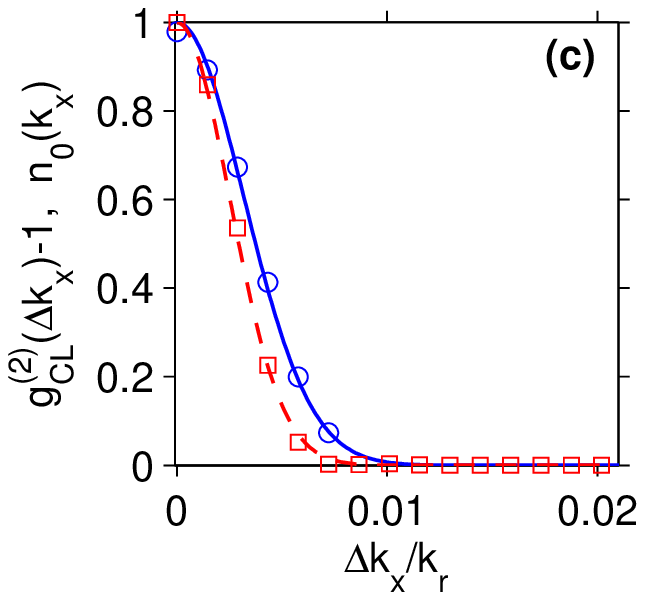}%
\includegraphics[height=5cm]{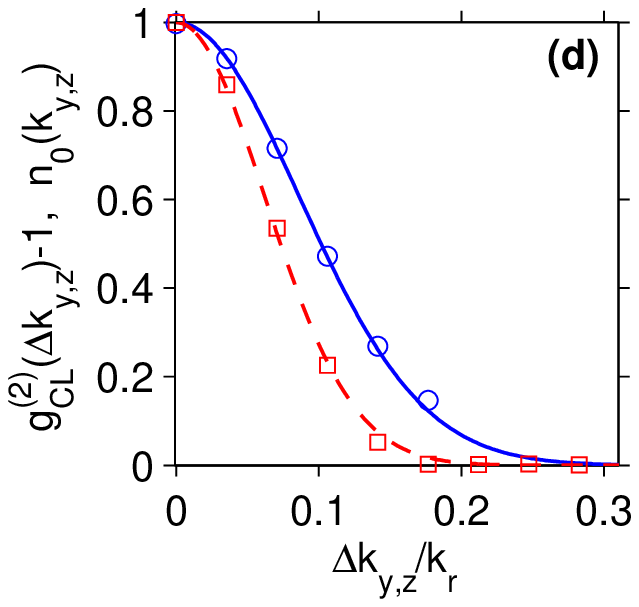}
\caption{Back-to-back (BB) and collinear (CL) atom-atom pair correlation, $%
g_{BB/CL}^{(2)}(\Delta k_{i})-1$ as a function of the displacement $\Delta
k_{i}$ ($i=x,y,z$) in units of the collision momentum $k_{r}$, after $t_{f}=25$ $%
\protect\mu $s collision time. The circles are the numerical
results, angle-averaged over the halo of scattered atoms
after elimination of the regions occupied by the two colliding
condensates; the solid lines are simple Gaussian fits to guide the
eye (see text). For comparison, we also plot the initial momentum
distribution $n_{0}(k_{i})$ of the colliding condensates; the actual
data points are shown by the squares and are fitted by a dashed-line
Gaussian.} \label{fig4}
\end{figure}

Figure~\ref{fig4} shows the numerical results for the back-to-back
and
collinear correlations (solid lines with circles), defined in Equations~(\ref%
{g2-CL-av}) and (\ref{g2-BB-av}). Due to the symmetry of the $y$ and
$z$ directions, the results in these directions are practically the
same. In order to verify the hypothesis that the shape and therefore
the width of the pair correlation functions is governed by the width
of the momentum distribution of the source condensate, we also plot
the actual initial momentum distributions of the source condensate
in the two orthogonal directions (with the understanding that the
horizontal axis $\Delta k_{i}$ now refers to the actual wave-vector
component $k_{i}$). The actual data points for the correlation
functions and for the momentum distribution of the source are shown
by the circles and squares, respectively, and are fitted with
Gaussian curves for simplicity and to guide the eye. The Gaussian
fits for the correlation functions
(solid lines) give:%
\begin{eqnarray}
g_{BB}^{(2)}(\Delta k_{i})-1 &=&9.2\exp \{-\Delta
k_{i}^{2}/[2(\sigma
_{i}^{BB})^{2}]\}, \\
g_{CL}^{(2)}(\Delta k_{i})-1 &=&\exp \{-\Delta k_{i}^{2}/[2(\sigma
_{i}^{CL})^{2}]\},
\end{eqnarray}
where the correlation widths $\sigma _{i}^{BB}$ and $\sigma
_{i}^{CL}$ are shown in the table~(\ref{corr_widths}) below.
The Gaussian fits (dashed lines) for the slices of the initial
momentum distribution $n_{0}(k_{i})\propto \exp
\{-k_{i}^{2}/[2(\sigma _{i})^{2}]\}$ are scaled to the same peak
value as $g_{BB/CL}^{(2)}(0)-1$ and have $\sigma _{x}=0.0025k_{r}$
and $\sigma _{y,z}=0.055k_{r}$.

By comparing the solid and the dashed lines we see that the shape of
the correlation functions indeed closely follow the shape of
momentum distribution of the source. More specifically, we find that
the following results provide the best fit to our numerical data:
\begin{equation}
\begin{tabular}{|c|c|c|c|}
\hline $\sigma _{x}^{BB}/\sigma_{x}$ & $\sigma _{y,z}^{BB}/\sigma_{y,z}$
& $\sigma _{x}^{CL}/\sigma_{x}$ & $\sigma _{y,z}^{CL}/\sigma_{yz}$ \\
\hline $1.18$ & $1.39$ & $1.27$ & $1.57$ \\ \hline
\end{tabular}
\label{corr_widths}
\end{equation}%
The ratios between the collinear and back-to-back correlation widths
are $\sigma _{x}^{CL}/\sigma _{x}^{BB}\simeq 1.08$ and $\sigma
_{y,z}^{CL}/\sigma _{y,z}^{BB}\simeq 1.13$.
The errors due to stochastic sampling on all quoted values of the
correlation widths are smaller than $3$\%.

The values for $\sigma _{y,z}^{CL}/\sigma_{y,z}$ and $\sigma
_{y,z}^{BB}/\sigma_{y,z}$ can be compared with the respective
experimentally measured values of table~(\ref{g2-expt-2}) and we see
reasonably good agreement, even though the numerical data are for a
much shorter collision time.
The remaining discrepancy between the numerical data at $t_{f}=25$
$\mu $s and the experimentally measured values after a $\sim 140$ $\mu
$s interaction time may be due to the evolution of the
condensates past $25$ $\mu $s, not attainable within the
positive-$P$ method.
The above numerical results for the correlation widths can be also
compared with the simple analytic estimate based on the Gaussian
Ansatz treatment of Equations~(\ref{corr_widths_gauss_CL}) and
(\ref{corr_widths_gauss_BB}). We find that the approximate analytic
results overestimate the back-to-back and collinear widths by $\sim
20\%$ and $40\%$, respectively, in the present example.

The amplitude of the correlation functions can also be inferred by
simple models. In fact, the collinear correlation function is a
manifestation of the Hanbury Brown and Twiss effect since it
involves pairs from two independent spontaneous scattering events
and we expect an amplitude of $g^{(2)}_{CL}(0)=2$
~\cite{Moelmer-Orsay}. This is in agreement with the positive-$P$
simulations. The back-to-back correlation amplitude, on the other
hand, can be substantially higher and display super-bunching
($g_{BB}^{(2)}(0)\gg 1$) \cite{Dissociation-Savage-etal-1,zin:06}
since the origin of this correlation is a simultaneous creation of a
pair of particles in a single scattering event.

In a simple qualitative model~\cite{Orsay-collision-experiment}, the
amplitude of the back-to-back correlation can be linked to the
inverse population of the atomic modes on the halo.
As we show in~\ref{sect:occupation-numbers}, this
model follows the trends observed in our first-principles numerical
simulations.

\subsection{Shorter collision time}

Here we present the results of numerical simulation for the same
parameters as in our main numerical example from Sec.
\ref{sect:main-example}, except that the data are analyzed at
$t_{f}=12.5$ $\mu $s, which is half the previous interaction time.
We found in Sec.~\ref{sect:main-example} that $\sigma_{yz}^{BB},
\sigma_{yz}^{CL}$ and the width of the halo $\delta k$ are all
nearly the same. In Sec. 3, however, we argue that the widths of the
correlation functions and the halo are governed by different limits
[Equations (\ref{corr_widths_gauss_BB}),(\ref{corr_widths_gauss_CL})
and (\ref{delta-k}) or (\ref{width-stim}), respectively]. The
example in this section illustrates this point.

\begin{figure}[tbp]
\centering\includegraphics[height=5cm]{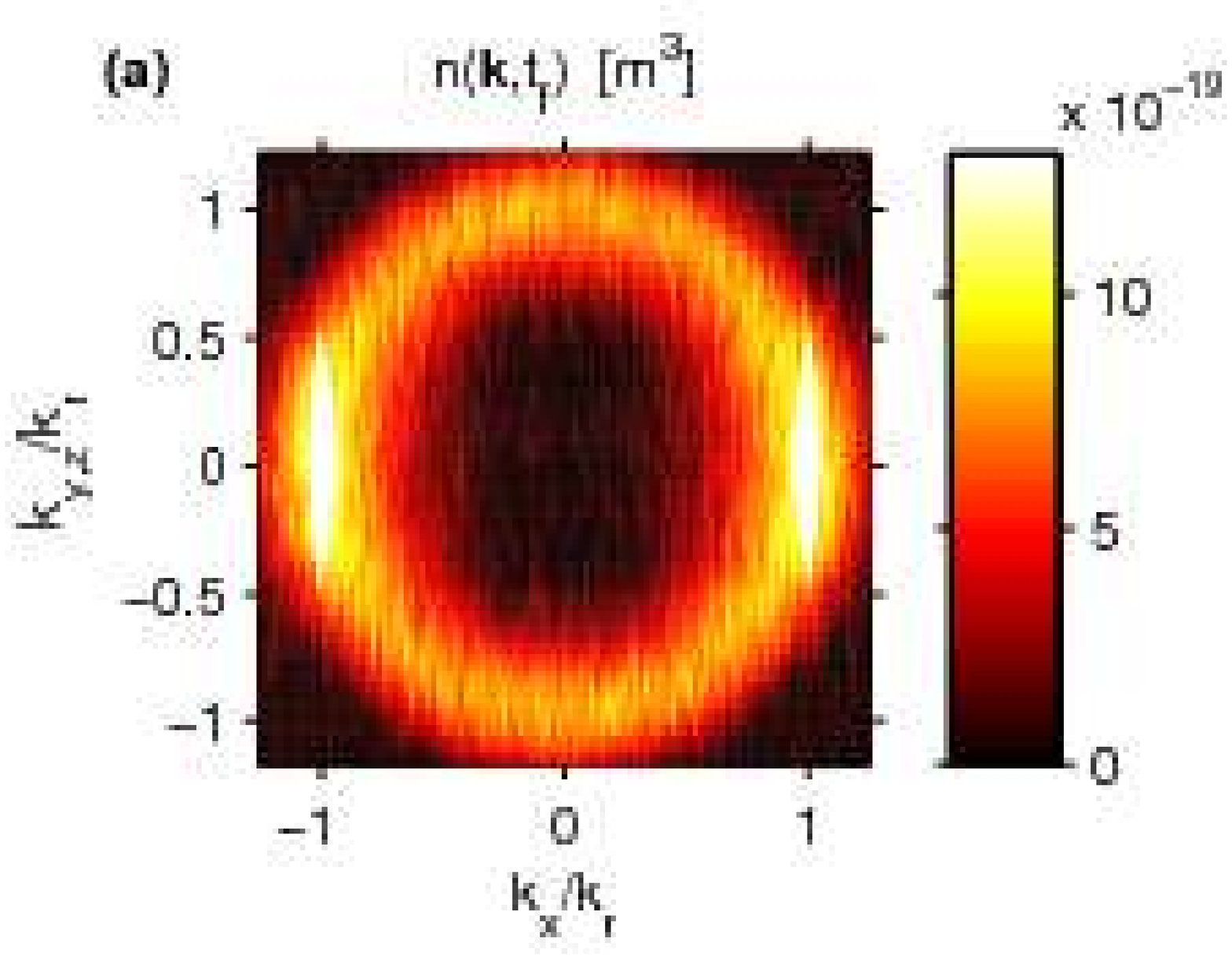}%
\includegraphics[height=5cm]{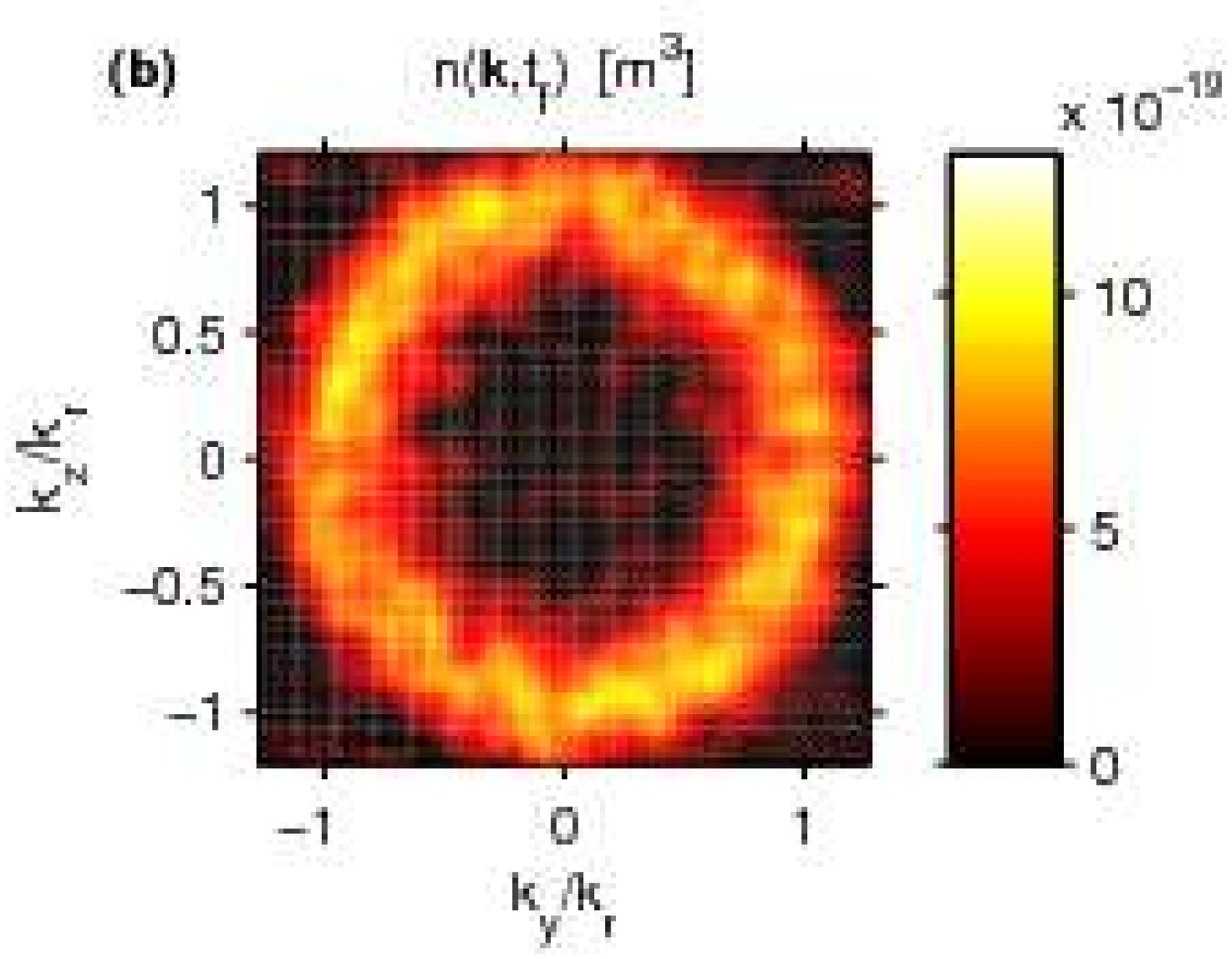}
\caption{Same as in figure~\protect\ref{fig2}, except for $t_{f}=12.5$ $%
\protect\mu $s collision time.}
\label{fig5}
\end{figure}

\begin{figure}[tbp]
\centering\includegraphics[height=5cm]{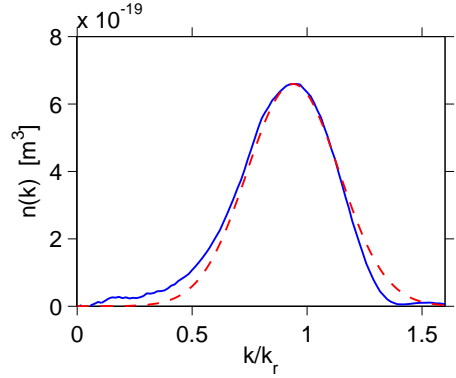}
\caption{Same as in figure~\protect\ref{fig3}, except for $t_{f}=12.5$ $%
\protect\mu $s collision time. The width and the peak of the fitted Gaussian
here are: $\delta k=0.20k_{r}$ and $k_{0}=0.95k_{r}$.}
\label{fig6}
\end{figure}

Figure~\ref{fig5} shows two orthogonal slices of the $s$-wave
scattering sphere in momentum space (cf. figure~\ref{fig2}), whereas
figure~\ref{fig6} is the corresponding radial distribution after
angle averaging. The most obvious feature of the distribution is
that it is broader than at $t_{f}=25$ $\mu$s and the fitted Gaussian
gives the radial width of $\delta k=0.20k_{r}$. This is precisely
twice the width in Figure~\ref{fig3} and is in agreement with the
simple qualitative estimate of \Eref{delta-k}.

The back-to-back and collinear correlation functions after
$t_{f}=12.5$ $\mu $s collision time are qualitatively very similar
to those shown in figure~\ref{fig4}, except that the Gaussian fits are
\begin{eqnarray}
g_{BB}^{(2)}(\Delta k_{i})-1 &=& 35.6\exp \{-\Delta
k_{i}^{2}/[2(\sigma
_{i}^{BB})^{2}]\}, \\
g_{CL}^{(2)}(\Delta k_{i})-1 &=& \exp \{-\Delta k_{i}^{2}/[2(\sigma
_{i}^{CL})^{2}]\},
\end{eqnarray}
with the correlation widths given by
\begin{equation}
\begin{tabular}{|c|c|c|c|}
\hline $\sigma _{x}^{BB}/\sigma_{x}$ & $\sigma _{y,z}^{BB}/\sigma_{y,z}$
& $\sigma _{x}^{CL}/\sigma_{x}$ & $\sigma _{y,z}^{CL}/\sigma_{yz}$\\
\hline $1.16$ & $1.28$ & $1.27$ & $1.48$ \\ \hline
\end{tabular}%
\end{equation}%
The ratios between the widths are $\sigma _{x}^{CL}/\sigma
_{x}^{BB}\simeq 1.09$ and $\sigma _{y,z}^{CL}/\sigma _{y,z}^{BB}
\simeq 1.16$.

For the correlation functions, the main difference compared to the
case for $25$ $\mu $s is that the peak
value of the back-to-back correlation is now larger, reflecting the
lower atomic density on the  scattering halo. The correlation
widths, on the other hand, are practically unchanged, at least within the
numerical sampling errors of the positive-$P$ simulations; the errors are at
the level of the third significant digit in the quoted values, which we
suppress. The number of scattered atoms in this example is about $850$,
which is approximately half the number at $25$ $\mu $s, confirming the
approximately linear dependence on time in the spontaneous scattering regime.

\subsection{Smaller collision velocity}

In this example, we present the results of simulations in which the
collision velocity is smaller by a factor $\sqrt{2}$ than before, $%
v_{r}^{\prime }=6.5$ cm/s ($k_{r}^{\prime }=4.1\times 10^{6}$ m$^{-1}$),
while all other parameters are unchanged.  In practice, this can be achieved
by changing the propagation directions of the Raman lasers that outcouple
the atoms from the trapped state. As in the previous example, the
halo width illustrates \Eref{delta-k}.

\begin{figure}[tbp]
\centering\includegraphics[height=5cm]{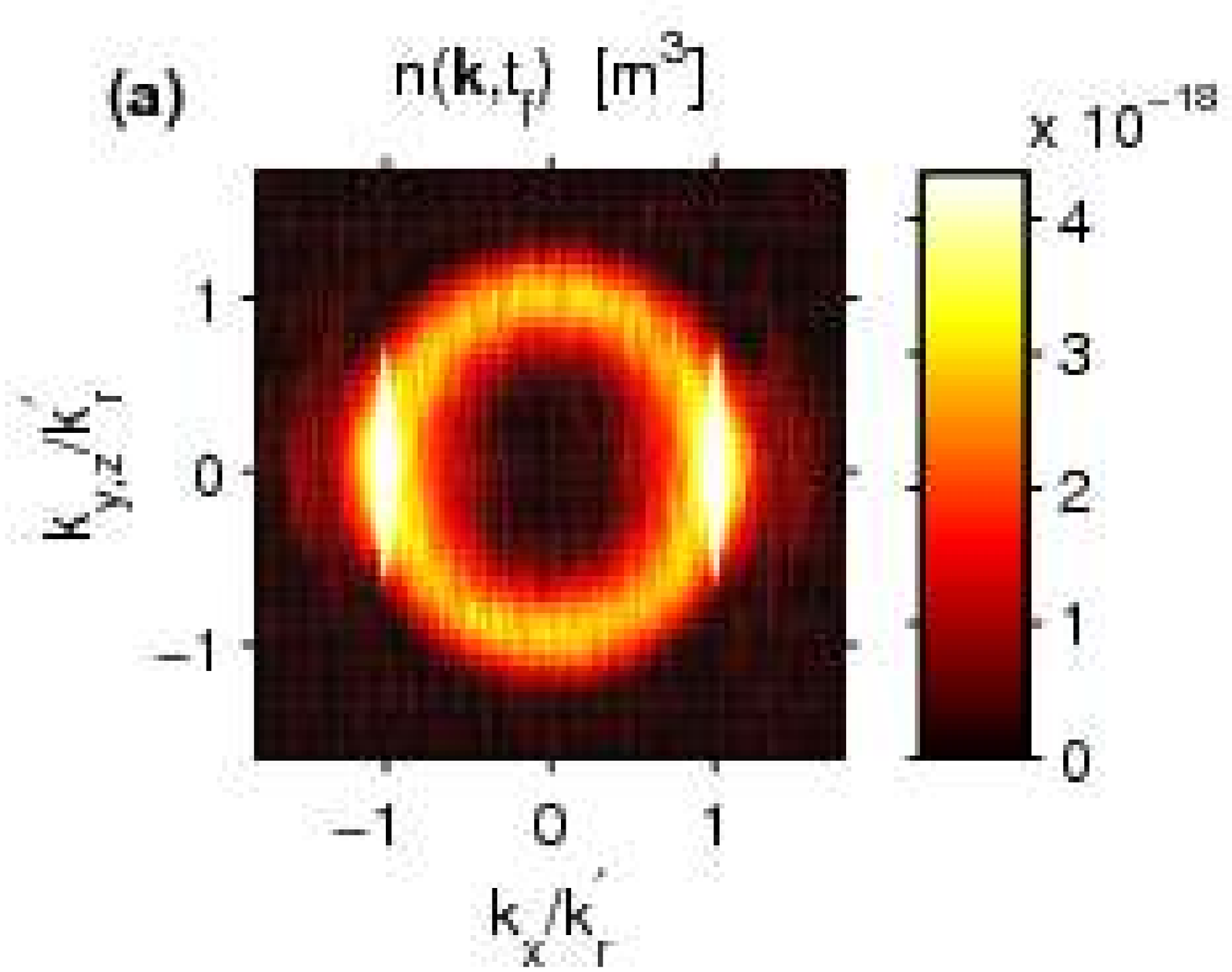}%
\includegraphics[height=5cm]{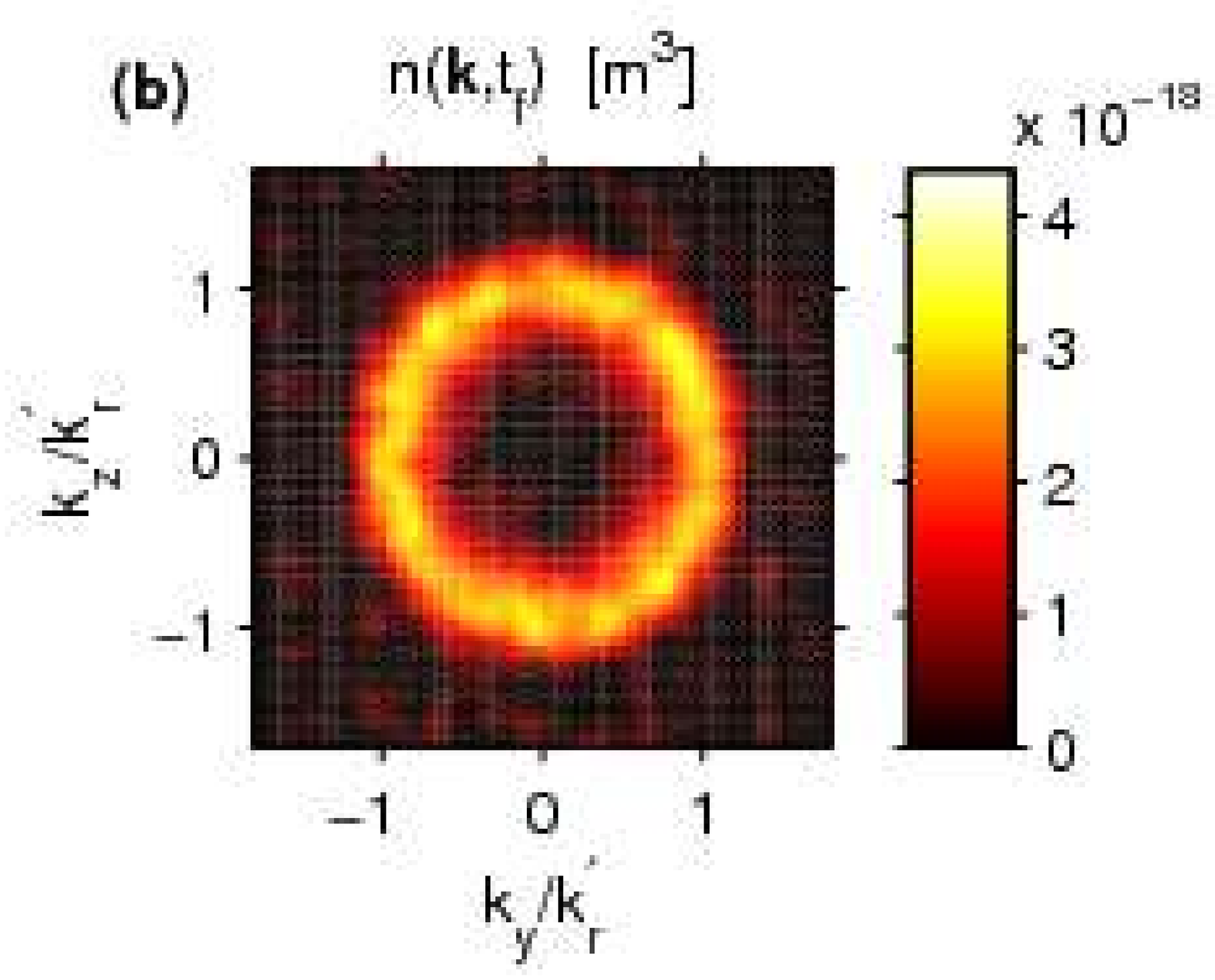}
\caption{Same as in figure~\protect\ref{fig2}, except for
$\protect\sqrt{2}$
times smaller collision velocity, $v_{r}^{\prime }=$ $6.46$ cm/s ($%
k_{r}^{\prime }=4.09\times 10^{6}$ m$^{-1}$). The axis for the momentum
components $k_{i}$ ($i=x,y,z$) are in units of the smaller recoil momentum
than in figure~\protect\ref{fig2}, and therefore the absolute radius of the $s$%
-wave scattering sphere is smaller in the present example. }
\label{fig7}
\end{figure}

\begin{figure}[tbp]
\centering\includegraphics[height=5cm]{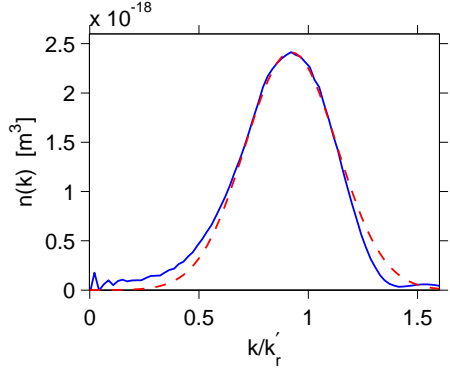}
\caption{Same as in figure~\protect\ref{fig3} except for
$\protect\sqrt{2}$ times smaller collision velocity $v_{r}^{\prime
}$ ($k_{r}^{\prime }=4.1\times 10^{6}$ m$^{-1}$). The width and the peak of the fitted Gaussian are $\delta k=0.21k_{r}^{\prime }=8.6\times 10^{5}$ m$^{-1}$; $k_{0}=0.92k_{r}^{\prime }$.} \label{fig8}
\end{figure}

The results of positive-$P$ simulations for the momentum density
distribution at $t_{f}=25$ $\mu $s are shown in figures~\ref{fig7} and
\ref{fig8}. The most obvious feature of the distribution is again
the fact that it is now broader than in our main example of Sec.~\ref%
{sect:main-example}. The width of the Gaussian function fitted to the
numerically calculated radial momentum distribution is given by $%
\delta k=0.21k_{r}^{\prime }$. This is again in excellent agreement with the
simple analytic estimate of \Eref{delta-k}, which predicts the
broadening to be inversely proportional to the collision velocity. We also
note that the peak momentum (relative to $k_{r}^{\prime }$) in the present
example is slightly shifted towards the centre of the halo, $%
k_{0}=0.92k_{r}^{\prime }$, which is a feature predicted in~\cite%
{Bach-Trippenbach-Rzazewski-2002} to occur when the ratio of the kinetic
energy to the interaction energy per particle is reduced.

The back-to-back and collinear correlation functions in this example
are again qualitatively very similar to those shown in
figure~\ref{fig4}, except that the Gaussian fits are
\begin{eqnarray}
g_{BB}^{(2)}(\Delta k_{i})-1 &=& 9.0\exp \{-\Delta
k_{i}^{2}/[2(\sigma
_{i}^{BB})^{2}]\}, \\
g_{CL}^{(2)}(\Delta k_{i})-1 &=& \exp \{-\Delta k_{i}^{2}/[2(\sigma
_{i}^{CL})^{2}]\},
\end{eqnarray}
with the correlation widths given by
\begin{equation}
\begin{tabular}{|c|c|c|c|}
\hline
$\sigma _{x}^{BB}/\sigma_{x}$ & $\sigma _{y,z}^{BB}/\sigma_{y,z}$ & $%
\sigma _{x}^{CL}/\sigma_{x}$ & $\sigma _{y,z}^{CL}/\sigma_{y,z}$
\\ \hline
$1.16$ & $1.35$ & $1.31$ & $1.51$ \\ \hline
\end{tabular}%
\end{equation}
where  $\sigma_{x}/k_{r}^{\prime }\simeq 0.0035$ and
$\sigma_{x}/k_{r}^{\prime }\simeq 0.078$. The ratios between the
collinear and back-to-back correlation widths are $\sigma
_{x}^{CL}/\sigma _{x}^{BB}\simeq 1.13$ and $\sigma
_{y,z}^{CL}/\sigma _{y,z}^{BB}\simeq 1.12$.

As we see from these results, the absolute widths of the correlation
functions are practically unchanged compared to the main numerical
example~(\ref{corr_widths}). This provides further evidence
that, at least for short collision times, the correlation widths are
governed by the momentum width of the source condensate, which is
unchanged in the present example compared to the case of Sec.
\ref{sect:main-example}.

The number of scattered atoms in this example is about $1270$, which is
approximately $\sqrt{2}$ times smaller than in Sec. \ref%
{sect:main-example} and corresponds to $\sim 1.3\%$ conversion. This
scaling is in agreement with the rate equation
approach~\cite{zin:06}, according to which the number of scattered
atoms is proportional to the square root of the collision energy and
hence to the collision momentum, which is $\sqrt{2}$ times smaller
here.

\subsection{Smaller scattering length}

Finally, we present the results of numerical simulations for the same
parameters as in our main numerical example from Sec.~\ref{sect:main-example}%
, except that the scattering lengths $a_{11}$ and $a_{00}$ are artificially
halved, i.e. $a_{00}=2.65$ nm and $a_{11}=3.75$ nm. The trap frequencies are
unchanged and we modify the chemical potential to arrive at the same peak
density of the initial BEC in the trap, $\rho _{0}\simeq 2.5\times 10^{19}$ m%
$^{-3}$. The total number of atoms is now
smaller, $N\simeq 3.5\times 10^{4}$.
One effect of changing the scattering length is that it changes the size and
shape of the trapped cloud, and therefore also its momentum distribution.
The shape is slightly closer to a Gaussian and therefore also to the
treatment in~\cite{Moelmer-Orsay}.

\begin{figure}[tbp]
\centering\includegraphics[height=5cm]{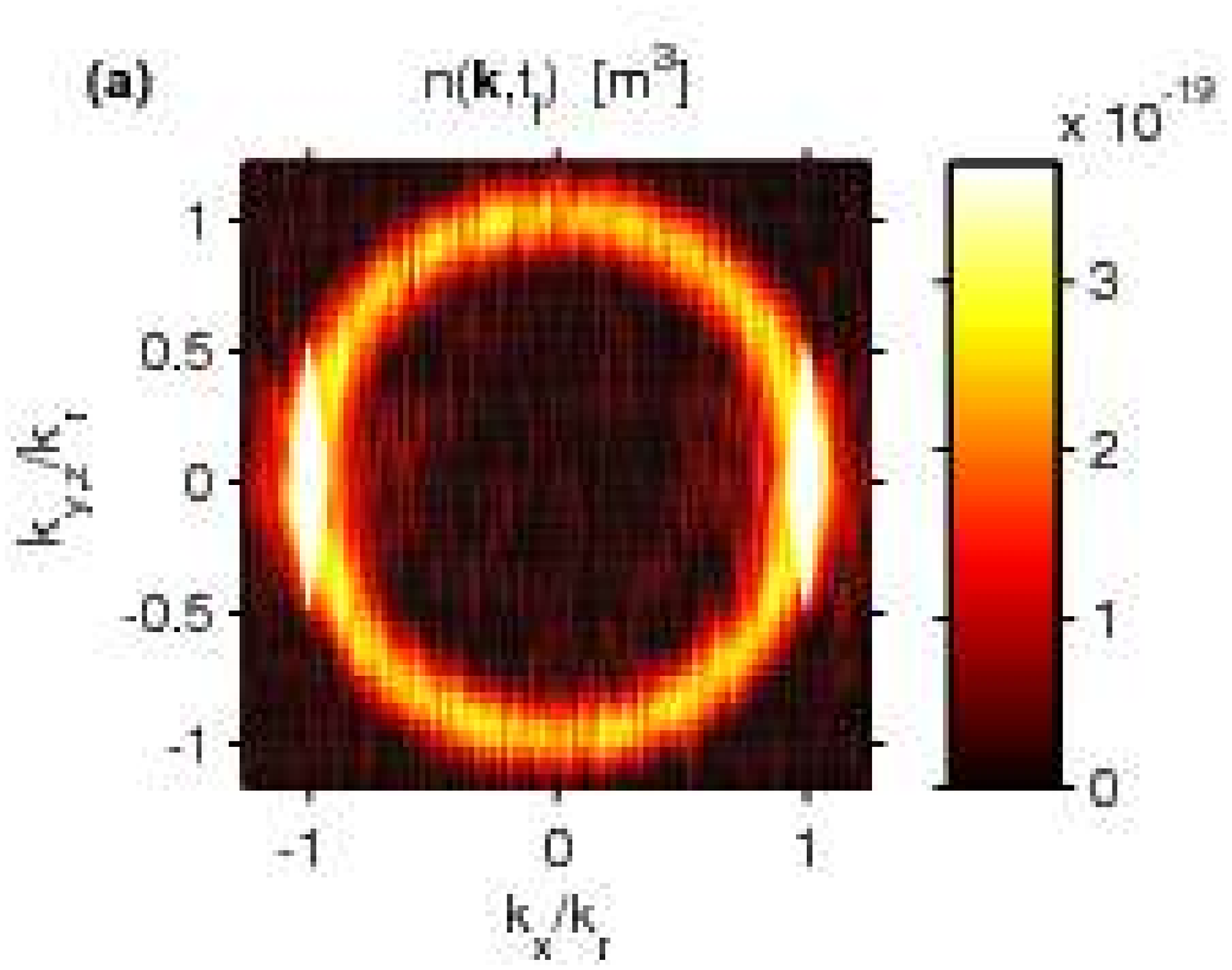} %
\includegraphics[height=5cm]{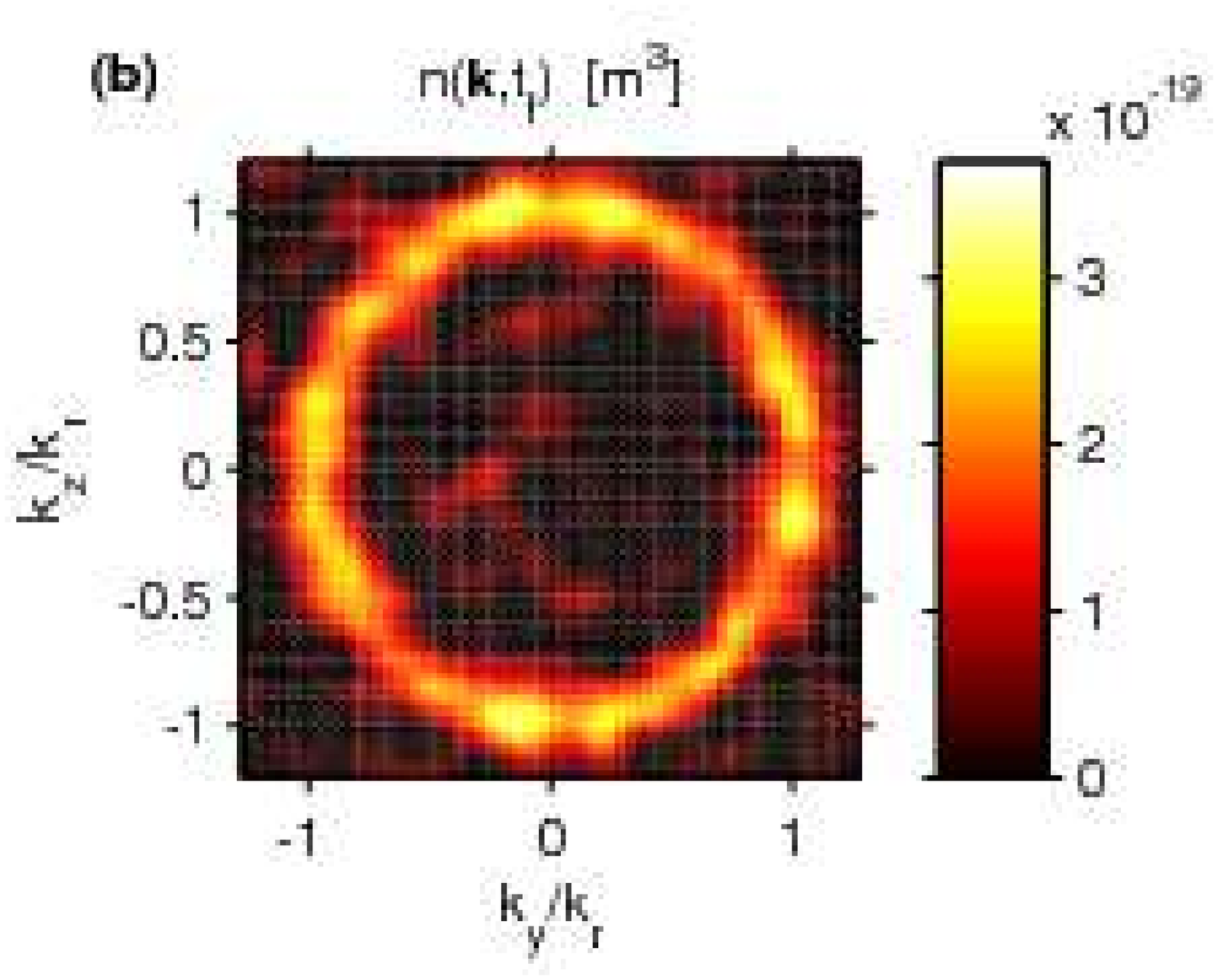}
\caption{Same as in figure~\protect\ref{fig2} except for the
scattering lengths of $a_{00}=2.65$ nm and $a_{11}=3.75$ nm, which
are twice smaller than before.} \label{fig9}
\end{figure}

\begin{figure}[tbp]
\centering\includegraphics[height=5cm]{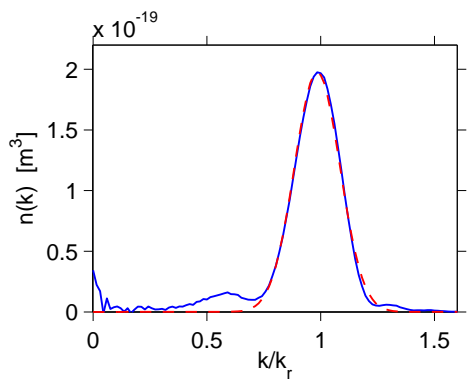}
\caption{Same as in figure~\protect\ref{fig3} except for twice smaller
values of the scattering lengthes $a_{00}$ and $a_{11}$. The width and the peak of the fitted Gaussian are $\delta k=0.10k_{r}$
and $k_{0}=0.98k_{r}$, which are the same as in
figure~\protect\ref{fig3}.}
\label{fig10}
\end{figure}

Due to the smaller scattering length, the density distribution in position
space of the initial trapped condensate is now narrower and conversely the
momentum distribution of the colliding condensates is broader. On the other
hand, the width of the halo (see figures~\ref{fig9} and
\ref{fig10} at $t_{f}=25$ $\mu $s) of scattered atoms is practically
unchanged compared to the example of figure~\ref{fig3}, as it is governed by
the energy-time uncertainty consideration~(\ref{delta-k}), for the
spontaneous scattering regime. The only quantitative difference is the lower
peak density on the scattering sphere, which is due to the weaker strength
of atom-atom interactions resulting in a slower scattering rate. The number
of scattered atoms at $25$ $\mu $s is $\sim 180$, corresponding to $0.51\%$
conversion of the initial total number $N\simeq 3.5\times 10^{4}$. The
fraction of $0.51\%$ itself corresponds approximately to a scaling law of $%
\sim a^{3/2}$, which is the same as the scaling of the total initial
number of trapped atoms in the Thomas-Fermi limit for a fixed peak density.

\begin{figure}[tbp]
\centering\includegraphics[height=5cm]{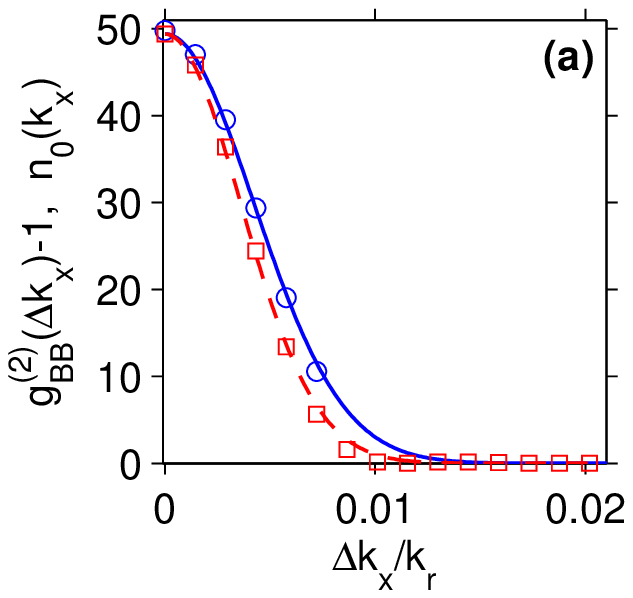} ~~~%
\includegraphics[height=5cm]{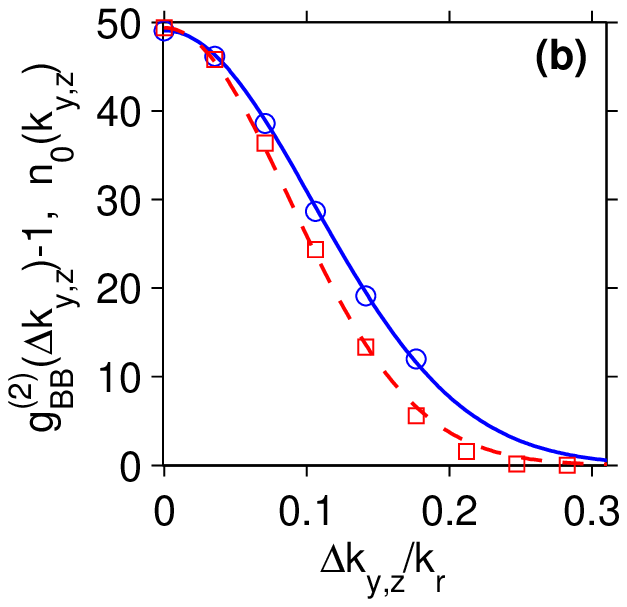} %
\includegraphics[height=5cm]{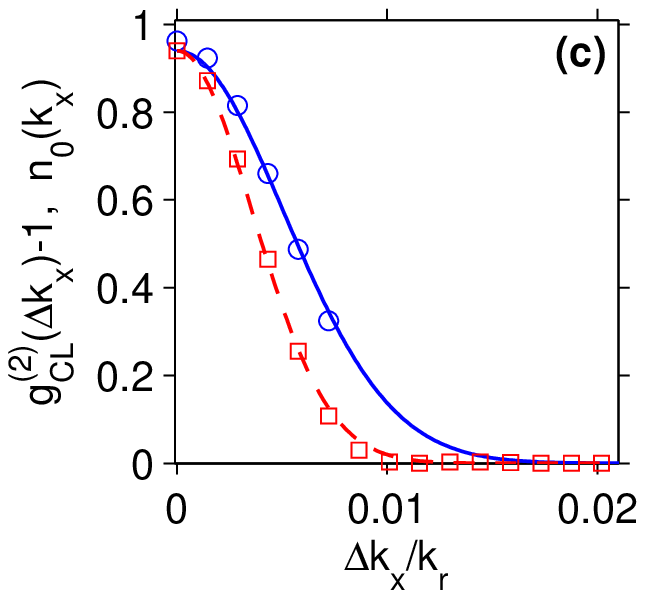} ~%
\includegraphics[height=5cm]{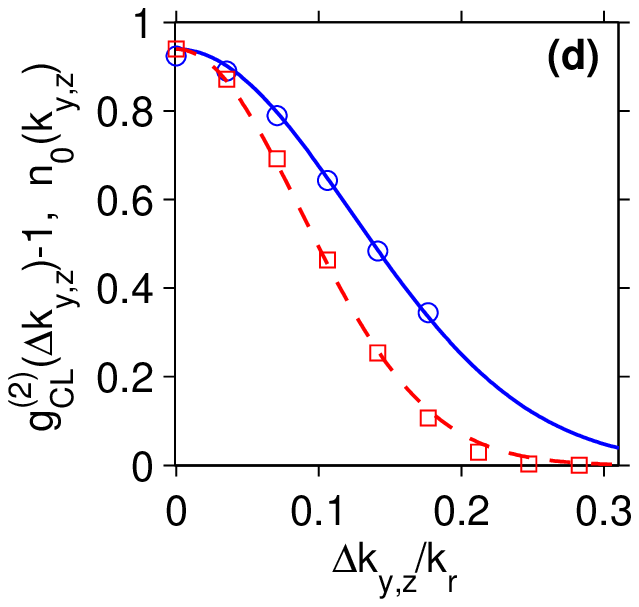}
\caption{Same as in figure~\protect\ref{fig4} except for twice smaller $s$%
-wave scattering lengths $a_{11}$ and $a_{00}$. } \label{fig11}
\end{figure}

Since the widths of the correlation functions are governed by the
width of the momentum distribution of the initial colliding
condensates, we expect corresponding broadening of the correlation
functions as well (see figure~\ref{fig11}). To quantify this effect,
we fit the momentum distribution of the initial BEC by a Gaussian
$\propto \exp \{-k_{i}^{2}/[2(\sigma _{i})^{2}]\}$, where
$\sigma_{x}=0.0036k_{r}$ and $\sigma_{y,z}=0.068k_{r}$ (cf. with
$\sigma _{x}=0.0025k_{r}$ and $\sigma_{y,z}=0.055k_{r}$ in
figure~\ref{fig4}, which are $\sim \sqrt{2}$ smaller). The Gaussian
fits to the correlation functions in figure~\ref{fig11} are
\begin{eqnarray}
g_{BB}^{(2)}(\Delta k_{i})-1 &=& 49\exp
\{\Delta k_{i}^{2}/[2(\protect\sigma _{i}^{BB})^{2}]\}, \\
g_{CL}^{(2)}(\Delta k_{i})-1 &=& 0.94\exp \{\Delta k_{i}^{2}/[2(\protect%
\sigma _{i}^{CL})^{2}]\},
\end{eqnarray}
where the widths $\protect\sigma _{i}^{BB}$ and $%
\protect\sigma _{i}^{CL}$ are given by
\begin{equation}
\begin{tabular}{|c|c|c|c|}
\hline $\sigma _{x}^{BB}/\sigma_{x}$ & $\sigma _{y,z}^{BB}/\sigma_{y,z}$
& $\sigma _{x}^{CL}/\sigma_{x}$ & $\sigma _{y,z}^{CL}/\sigma_{y,z}$ \\
\hline $1.18$ & $1.53$ & $1.42$ & $1.81$
\\ \hline
\end{tabular}
\label{widths-small-a}
\end{equation}

We see that the relative widths are practically unchanged, implying
that the absolute widths are broadened. The ratios between the
collinear and back-to-back correlation widths are slightly increased
and are given by $\sigma _{x}^{CL}/\sigma _{x}^{BB}\simeq 1.20$ and
$\sigma _{y,z}^{CL}/\sigma _{y,z}^{BB}\simeq 1.18$.

These numerical results make the present example -- with the
diminished role of atom-atom interactions -- somewhat closer to the
simple analytic
predictions of Equations~(\ref{corr_widths_gauss_CL}) and (\ref%
{corr_widths_gauss_BB}) based on a Gaussian ansatz for noninteracting
condensates.

\subsection{Relative number squeezing and violation of Cauchy-Schwartz
inequality}

Another useful measure of atom-atom correlations is the normalized variance
of the relative number fluctuations between atom numbers $\hat{N}_{i}$ and $%
\hat{N}_{j}$ in a pair of counting volume elements denoted via $i$ and $j$,%
\begin{equation}
V_{i-j}=\frac{\langle \lbrack \Delta (\hat{N}_{i}-\hat{N}_{j})]^{2}\rangle }{%
\langle \hat{N}_{i}\rangle +\langle \hat{N}_{j}\rangle }=1+\frac{\langle
:[\Delta (\hat{N}_{i}-\hat{N}_{j})]^{2}:\rangle }{\langle \hat{N}_{i}\rangle
+\langle \hat{N}_{j}\rangle },  \label{variance}
\end{equation}%
where $\Delta \hat{X}=\hat{X}-\langle \hat{X}\rangle $ is the fluctuation.
This definition uses the conventional normalization with respect to the
shot-noise level characteristic of Poissonian statistics, such as for a
coherent state, $\langle \hat{N}_{i}\rangle +\langle \hat{N}_{i}\rangle $.
In this case the variance $V_{i-j}=1$, which corresponds to the level of
fluctuations in the absence of any correlation between $\hat{N}_{i}$ and $%
\hat{N}_{j}$. Variance smaller than one, $V_{i-j}<1$, implies reduction (or
squeezing) of fluctuations below the shot-noise level and is due to quantum
correlation between the particle number fluctuations in $\hat{N}_{i}$ and $%
\hat{N}_{j}$. Perfect ($100$\%) squeezing of the relative number
fluctuations corresponds to $V_{i-j}=0$.

\begin{figure}[tbp]
\centering\includegraphics[height=5cm]{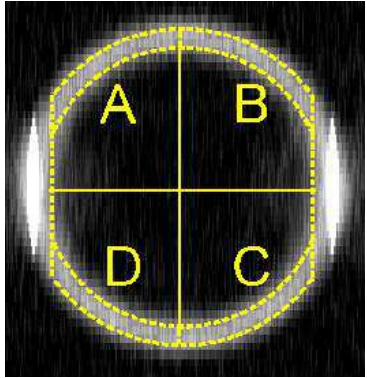}
\caption{Illustration of the four regions of the momentum space
density, forming the quadrants $A$, $B$, $C$, and $D$ on the
$s$-wave scattering sphere, on which we analyse the data for
relative number squeezing.} \label{fig12}
\end{figure}

In the context of the present model for the BEC collision experiment
and possible correlation measurements between atom number
fluctuations on diametrically opposite sides of the $s$-wave
scattering sphere, we assign the indices $i,j=A,B,C,D$ in \Eref{variance} to one of the four quadrants as illustrated in
figure~\ref{fig12}. The total atom number
operator $\hat{N}_{i}$ in each quadrant $\mathcal{D}_{i}$ within the $s$%
-wave scattering sphere is defined after elimination of the regions in
momentum space occupied by the two colliding condensates%
\begin{equation}
\hat{N}_{i}(t)=\int_{\mathcal{D}_{i}}dk_{x}dk_{y}\int\nolimits_{-\infty
}^{+\infty }dk_{z}\hat{n}(\mathbf{k},t).
\end{equation}%
Operationally, this is implemented by discarding the data points beyond $%
|k_{x}|>0.8k_{r}$. In addition, the quadrants $\mathcal{D}_{i}$ are defined
on a 2D plane after integrating the momentum distribution along the $z$%
-direction, which in turn only takes into account the 3D data points
satisfying $|1-k^{2}/k_{r}^{2}|<0.28$, i.e. lying in the narrow
spherical shell $k_{r}\pm \delta k$ with $\delta k\simeq 0.14k_{r}$.
The elimination of the inner and outer regions of the halo is done
to minimize the sampling error in our simulations, since these
regions have vanishingly small population and produce large noise in
the stochastic simulations.

\begin{figure}[tbp]
\centering\includegraphics[height=9cm]{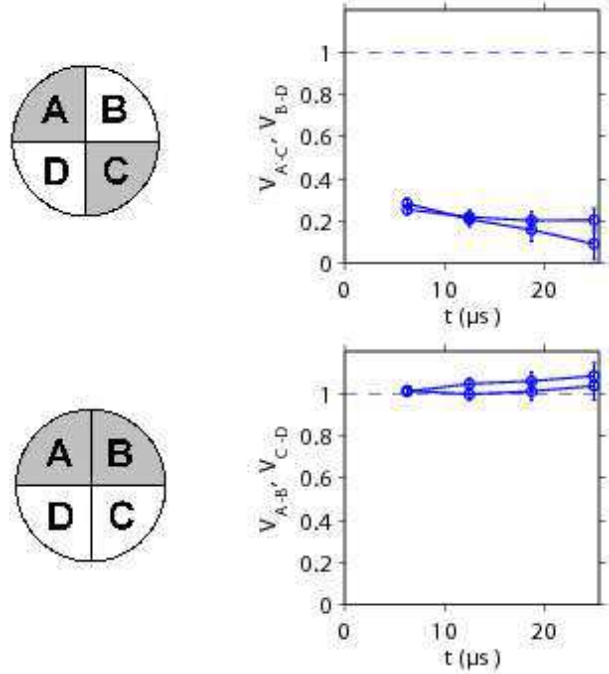}
\caption{Relative number variance in the diametrically opposite and
neighboring quadrants, $V_{A-C/B-D}$ and $V_{A-B/C-D}$, as a
function of time.} \label{fig13}
\end{figure}

The choice of the quadrants as above is a particular implementation of the
procedure of binning, known to result in a stronger correlation signal and
larger relative number squeezing \cite%
{Dissociation-Greiner,Dissociation-Savage-etal-2}. Due to strong
back-to-back pair correlations, we expect the relative number fluctuations
in the diametrically opposite quadrants to be squeezed, $V_{A-C},V_{B-D}<1$,
while the relative number variance in the neighboring quadrants, such as $%
V_{A-B}$ and $V_{C-D}$, is expected to be larger than, or equal to,
one. The positive-$P$ simulations confirm these expectations and are
shown in figure~\ref{fig13}, where we see strong ($\sim 80\%$)
relative number squeezing for the diametrically opposite quadrants,
$V_{A-C,B-D}\simeq 1-0.8=0.2$.

These results assume a uniform detection efficiency of $\eta=1$,
whereas if the efficiency is less than $100$\% ($\eta<1$), then the
second term in \Eref{variance} should be multiplied by $\eta$.
This implies, that for $\eta=0.1$ as an example, the above
prediction of $\sim 80$\% relative number squeezing will be degraded
down to a much smaller but still measurable value of $\sim 8$\% squeezing
($V_{A-C,B-D}\simeq 1-0.08=0.92$). Even with perfect
detection efficiency, our simulations do not lead to ideal ($100$\%)
squeezing. This can be understood in terms of a small fraction of
collisions that take place with a center-of-mass momentum offset
that is (nearly) parallel to one of the borders between the
quadrants. As a result, the respective scattered pairs fail to
appear in diametrically opposite quadrants during the (finite)
propagation time (see also~\cite{Dissociation-Savage-etal-2}).

For the symmetric case with $\langle \hat{N}_{i}\rangle =\langle
\hat{N}_{j}\rangle $ and $\langle \hat{N}_{i}^{2}\rangle =\langle
\hat{N}_{j}^{2}\rangle $, the variance $V_{i-j}$ can be rewritten as
\begin{equation}
V_{i-j}=1+\langle \hat{N}_{i}\rangle \lbrack g_{ii}^{(2)}-g_{ij}^{(2)}],
\label{V-alt}
\end{equation}%
where the second-order correlation function $g_{ij}^{(2)}$ is defined
according to%
\begin{equation}
g_{ij}^{(2)}=\frac{\langle :\hat{N}_{i}\hat{N}_{j}:\rangle }{\langle \hat{N}%
_{i}\rangle \langle \hat{N}_{j}\rangle }.  \label{g2-bar}
\end{equation}

Equation (\ref{V-alt}) helps to relate the relative number squeezing, $%
V_{i-j}<1$, to the violation of the classical Cauchy-Schwartz inequality $%
g_{12}^{(2)}>g_{11}^{(2)}$, studied extensively in quantum optics
with photons \cite{Zou-Mandel,Walls-Milburn}. The analysis presented
here (see also~\cite{Dissociation-Savage-etal-2} on molecular
dissociation) shows that the Cauchy-Schwartz inequality, and its
violation, is a promising area of study in \textit{quantum atom
optics} as well.

\section{Summary}

An important conclusion that we can draw from the numerical simulations is
that the predicted widths of the correlation functions
are remarkably robust against the parameter variations
we were able to explore  (in Secs. 5.1 through 4).  This gives us confidence in
our physical interpretation of the width as being chiefly due to the initial
momentum width of the condensate. The discrepancy with the analytical calculation of~\cite{Moelmer-Orsay} seems to be primarily due to the different cloud shapes used.
The width of the halo varies with the parameters we tested in a predictable way and also
confirms the discussion in Sec. \ref{qualitative-analysis}.

As for comparison with the experiment, the numerically calculated widths of the scattering halo and the correlation functions
coincide with the experimental ones to within better than 20\% in most cases.
The main discrepancy with the experiment is in the \emph{ratio} of the back to back and collinear correlation widths.
From the experimental point of view, these ratios are more significant than the individual
widths since some sources of uncertainty,
such as the number of atoms and the size of the condensates, cancel.
The discrepancy may mean that the collinear correlations are not sufficient
to characterize the size and momentum distribution in the source at this level of accuracy.
The discrepancies may of course also be due to the numerous experimental imperfections,
especially the fact that the Raman outcoupling was only 60\% efficient, and therefore an
appreciable trapped $m_x=1$ condensate was left behind.
This defect may be remedied in future experiments.
On the other hand, the current simulations neglect the unavoidable interaction of the scattered
atoms with unscattered, $m_x=0$ condensates as they leave the interaction region.
This interaction could alter the trajectories of
the scattered atoms in a minor, but complicated way.
Future numerical work must examine this possibility further.

Still, the overall message of this work is that a first principles quantum field theory approach
can quantitatively account for experimental observations of atomic four-wave mixing
experiments.
This work represents the first time that this sort of numerical simulation has been carefully confronted
with an experiment.
An interesting extension would be to examine the regime of stimulated scattering.
It has been predicted that a highly anisotropic  BEC could lead to an anisotropic population of the scattering  halo~\cite{vardi:02,pu:00}. This effect would be a kind of atomic analogue of superradiance observed when off-resonant light is shined on a condensate~\cite{inouye:99, gross:82}.
In addition, our results may be useful beyond the cold atom community:
theoretical descriptions of correlation measurements in heavy ion collisions~\cite{wong:07}
may benefit from some of our insights.

\ack


The authors acknowledge stimulating discussions with A. Aspect, K. M{\o}lmer, M.
Trippenbach, P. Deuar, and M. Davis, and thank the developers of the XMDS
software \cite{xmds}. CS and KK acklowledge support from the Australian
Research Council and the Queensland State Government. Part of this work was
done at the Institut Henri Poincare -- Centre Emile Borel; KK thanks the
institute for hospitality and support. The atom optics group is supported
by the SCALA program of the EU, and by the IFRAF institute.

\appendix

\section{Duration of the collision}

\label{sect:collisionduration}

In order to estimate the collision duration one can consider a simple
classical model of the collision~\cite{zin:06}. Denoting by $\rho _{1}(%
\mathbf{x},t)$ and $\rho _{2}(\mathbf{x},t)$ the density distributions of
the two condensates, the number of scattered atoms $N_{\mathrm{sc}}(t)$ at a given time can be written
\begin{equation}
N_{\mathrm{sc}}(t)=2\int_{0}^{t}dt^{\prime }\int d^{3}\mathbf{x}%
~2\sigma _{0}v_{r}\rho _{1}(\mathbf{x},t^{\prime })\rho _{2}(\mathbf{x}%
,t^{\prime })  \label{eq:coltime}
\end{equation}%
where $\sigma _{0}=8\pi a_{00}^2$ is the cross section for a collision of two
particles. In this latter formula $a_{00}\simeq 5.3$ nm is the scattering
length between $m_{x}=0$ atoms~\cite{Orsay-collision-experiment}.

The time-dependent density of the two condensates can be calculated from the
expansion of a condensate in the Thomas-Fermi regime described in~\cite%
{castin:96}. This approach suggests two different time scales for
the collision duration. First, the separation of the two condensates
occurs in a time defined by the ratio of the longitudinal size of
the condensates and their relative velocity
$t_{\mathrm{sep}}=R_{x}/v_{r}$. Taking for $R_{x}$
the Thomas-Fermi radius of the initial condensate, one can show that $t_{%
\mathrm{sep}}$ is on the order of $1$~ms. At the same time, the
condensates expand during their separation on a time scale
$t_{\mathrm{exp}}=1/\omega _{y}=1/\omega _{z}\simeq 140~\mu $s. This
latter effect appears to be
predominant in the evaluation of \Eref{eq:coltime} and $t_{\mathrm{exp}%
} $ can be taken as a definition of the collision duration $\Delta t$. The numerical evaluation of \Eref{eq:coltime} gives $%
N_{\mathrm{sc}}(\Delta t)\simeq 0.66N_{\mathrm{sc}}(\infty )$
and the estimated total number of scattered atoms corresponds to the experimentally
observed $5\%$ of the initial total number of atoms in the trapped
condensate.

\section{Occupation number of the scattering modes and amplitude of the back-to-back correlation}

\label{sect:occupation-numbers}

In order to estimate the occupation number of the scattering modes
one needs to compare the number of scattered atoms $N_{sc}$ to the
number of scattering modes $N_{m}$. To achieve this one has to first
consider the volume of a scattering mode $V_{m}$, given by the
first-order coherence volume (also dubbed as \textquotedblleft phase
grain\textquotedblright  in~\cite{norrie:040401,Deuar-Drummond-4WM}).
Such a volume corresponds in
fact to the coherence volume of the source condensate, and in
practice it can also be deduced from the measurement of the width of
the collinear correlation function $g_{CL}^{(2)}(\Delta k_{i})$ as
one expects in a Hanbury Brown-Twiss experiment. For simplicity
we match the scattering mode volume $V_{m}$ to the coherence volume
of the source condensate in momentum space,
\begin{equation}
V_{m}\simeq \beta \sigma_{x}(\sigma_{yz})^{2},
\end{equation}%
where $\beta $ is a geometrical factor which depends on the geometry
of the modes. Approximating the source condensate in momentum space
by a Gaussian $\propto \exp
[-x^{2}/(2\sigma_{x}^{2})-(y^{2}+z^{2})/(2\sigma_{y,z}^{2})]$, one
has $\beta =(2\pi )^{3/2}$.

The number of scattering modes $N_{m}$ can in turn be estimated from
the knowledge of the total volume of the scattering shell $V$,
\begin{equation}
N_{m}=\frac{V}{V_{m}},
\end{equation}%
where the volume $V$ is determined from the value of the width of the
scattering shell $\delta k$:
\begin{eqnarray}
V&=&\int d^{3}\mathbf{k}~\exp [-(k-k_{r})^{2}/(2\delta k^{2})]\\
&\simeq& 4\pi\sqrt{2\pi}k_r^{2}\delta k,
\end{eqnarray}%
for $\delta k\ll k_r$. If we apply this estimate to the results of the
main numerical example (see Sec. \ref{sect:main-example}), we find
$N_{m}\simeq 26400$. As $N_{sc}=1750$, this implies an occupation
number per mode of $N_{sc}/N_{m}\simeq 0.066$. Such an estimate
confirms that the system is indeed in the spontaneous regime and
that bosonic stimulation effects are negligible.

The simple model of~\cite{Orsay-collision-experiment} for the
back-to-back correlation predicts that its height is given by
\begin{equation}%
g^{(2)}_{BB}(0)=1+N_{m}/N_{sc}
\end{equation}%
Using the above estimate of $N_{m}$ and the actual value of $N_{sc}$
found from the numerical simulations, we obtain that the height of
the back-to-back correlation peak should be approximately given by
$\sim 16$. This compares favorably with the actual numerical result
of $10.2$. Similarly, we obtain the back-to-back correlation peak
of: $\sim 62$ in the example with the shorter collision time
(compare with the numerical result of $36.6$); $\sim 18$ in the
example with the smaller collision velocity (compare with $10$); and
$\sim 70$ in the example with the smaller scattering length (compare
with $50$).

\section{Width of the $s$-wave scattering sphere in the undepleted
\textquotedblleft pump\textquotedblright\ approximation}

\label{sect:undepletedpump}

To estimate the width of the halo of scattered atoms beyond the
spontaneous regime we use the analytic solutions for a uniform system in the
so called undepleted \textquotedblleft pump\textquotedblright\ approximation
in which the number of atoms in the colliding condensates are assumed constant. This approximation
is applicable to short collision times. Nevertheless, it formally describes
the regime of stimulated scattering and can be used to estimate the width of
the $s$-wave scattering sphere as we show here.

The problem of BEC collisions in the undepleted \textquotedblleft
pump\textquotedblright\ approximation was studied in~\cite%
{Bach-Trippenbach-Rzazewski-2002}; the solutions for the momentum
distribution of the $s$-wave scattered atoms are formally equivalent to
those obtained for dissociation of a BEC of molecular dimers in the
undepleted molecular condensate approximation \cite%
{Dissociation-Savage-etal-1,Dissociation-Twin-Beams}. For a uniform
system with periodic boundary conditions, one has the following
analytic solution for momentum mode
occupation numbers:%
\begin{equation}
n_{\mathbf{k}}(t)=\frac{\overline{g}^{2}}{\overline{g}^{2}-\Delta _{k}^{2}}%
\sinh ^{2}\left( \sqrt{\overline{g}^{2}-\Delta _{k}^{2}}\,t\right) .
\label{n-k-analytic-b}
\end{equation}%
Here, the constant $\overline{g}$ is given by
\begin{equation}
\overline{g}=2U_{0}\rho_{0}=\frac{8\pi \hbar a_{00}\rho_{0}}{m},
\end{equation}%
where $U_{0}=4\pi\hbar a_{00}/m$ corresponds to the coupling constant $%
g/\hbar $ of~\cite{Bach-Trippenbach-Rzazewski-2002}, and we note that
the results of~\cite{Bach-Trippenbach-Rzazewski-2002} contain
typographical errors and have to be corrected as follows~\cite%
{Trippenbach-private}: given the Hamiltonian of~(1), with $g=4\pi
\hbar^{2} a/m$, the coupling $g$ in~(2), (7), (9), and (10), as well as
in the definition of $\Delta (p)$ after~(9), should be replaced by $2g$.
In the problem of molecular dissociation, the constant $\overline{g}$
corresponds to $\overline{g}=\chi \sqrt{\rho_{0}}$ \cite%
{Dissociation-Savage-etal-1}, where $\chi $ is the atom-molecule coupling
and $\rho_{0}$ is the molecular BEC density.

The parameter $\Delta _{k}$ in \Eref{n-k-analytic-b} corresponds to the
energy offset from the resonance condition%
\begin{equation}
\hbar \Delta _{k}\equiv \frac{\hbar ^{2}k^{2}}{2m}-\frac{\hbar ^{2}k_{r}^{2}%
}{2m},\label{eq:resonance}
\end{equation}%
where $\hbar k_{r}$ is the collision momentum; in molecular dissociation, $%
\hbar ^{2}k_{r}^{2}/m$ corresponds to the effective dissociation energy $%
2\hbar |\Delta _{eff}|$, using the notations of~\cite%
{Dissociation-Savage-etal-1}.

From \Eref{n-k-analytic-b} we see that modes with $\overline{g}%
^{2}-\Delta _{k}^{2}>0$ experience Bose enhancement and grow exponentially
with time, whereas the modes with $\overline{g}^{2}-\Delta _{k}^{2}<0$
oscillate at the spontaneous noise level. The absolute momenta of the
exponentially growing modes lie near the resonant momentum $\hbar k_{r}$,
and therefore we can use the condition $\overline{g}^{2}-\Delta _{k}^{2}=0$
to define the approximate width of the $s$-wave scattering sphere. First we
write $k=k_{r}+\Delta k$ and assume for simplicity that $k_{r}$ is large
enough so that $\Delta k\ll k_{r}$. Then the condition $\overline{g}%
^{2}-\Delta _{k}^{2}=0$ can be approximated by
\begin{equation}
1-\left( \frac{\hbar k_{r}\Delta k}{m\overline{g}}\right) ^{2}\simeq 0.
\end{equation}%
This can be solved for $\Delta k$ and used to define the width $\delta
k=\Delta k/2$ of the $s$-wave scattering sphere as
\begin{equation}
\frac{\delta k}{k_{r}}\simeq \frac{m\overline{g}}{2\hbar k_{r}^{2}}=\frac{%
4\pi a_{00}\rho_{0}}{k_{r}^{2}}.  \label{width}
\end{equation}%
The reason for defining it as half of $\Delta k$ is to make $\delta
k$ closer in definition to the half-width at half maximum and to the
rms width around $k_{r}$.

The above simple analytic estimate (\ref{width}) gives $\delta
k/k_{r}\simeq 0.05$ for the present $^{4}$He$^{*}$ parameters. For
comparison, the actual width of the analytic result
(\ref{n-k-analytic-b}) varies between $\delta k/k_{r}\simeq 0.12$
and $\delta k/k_{r}\simeq 0.027$ for durations between
$\overline{g}t=1$ and $\overline{g}t=7$, corresponding,
respectively, to $t\simeq 20$ $\mu$s and  $t\simeq 140$ $\mu$s in
the present $^{4}$He$^{*}$ example.

\section{Positive-$P$ simulation parameters}

\label{sect:parameters}

The positive-$P$ simulations in our main numerical example of Sec. \ref%
{Numerical results} are performed on a computational lattice with $%
1400\times 50\times 70$ points in the ($x,y,z$)-directions
respectively. The
length of the quantization box along each dimension is $L_{x}=252$ $\mu $m, $%
L_{y}=20.52$ $\mu $m, and $L_{z}=30.76$ $\mu $m. The computational lattice
in momentum space is reciprocal to the position space lattice and has the
lattice spacing of $\Delta k_{i}=2\pi /L_{i}$, giving $\Delta
k_{x}=2.49\times 10^{4}$ m$^{-1}$, $\Delta k_{y}=3.06\times 10^{5}$ m$^{-1}$%
, and $\Delta k_{z}=2.04\times 10^{5}$ m$^{-1}$. The momentum cutoffs are $%
k_{x}^{(\max )}=1.75\times 10^{7}$ m$^{-1}$, $k_{y}^{(\max )}=7.66\times
10^{6}$ m$^{-1}$, and $k_{z}^{(\max )}=7.15\times 10^{6}$ m$^{-1}$.

The momentum cutoff in the collision direction, $k_{x}^{(\max )}$,
is more than $3$ times larger than the collision momentum $k_{r}$,
and hence it captures all relevant scattering processes of interest,
including the energy non-conserving scatterings
$(k_{r})+(k_{r})\rightarrow (3k_{r})+(-k_{r})$ and
$(-k_{r})+(-k_{r})\rightarrow (-3k_{r})+(k_{r})$
\cite{Deuar-Drummond-4WM}. In all our figures, the regions of
momentum space covering $k_{x}\simeq \pm 3k_{r}$ are not shown for
the clarity of presentation of the main halo. These scattering
processes, which produce a weak but not negligible signal at
$k_{x}\simeq \pm 3k_{r}$, i.e., outside the main halo are enhanced
by Bose stimulation due to the large population of the colliding
condensate components at $k_{x}\simeq \mp k_{r}$, respectively. In
the remaining $y$ and $z$ directions, such processes are absent and
therefore the number of lattice points and the momentum cutoffs can
be smaller.

Since the momentum distribution of the initial condensate is the narrowest
in the $k_{x}$-direction, one may question whether the resolution of $\Delta
k_{x}=2.49\times 10^{4}$ m$^{-1}$ with $1400$ lattice points is sufficient.
We check this by repeating the simulations with $4200\times 40\times 40$
lattice points and quantization lengths of $L_{x}=753$ $\mu $m and $%
L_{y}=L_{z}=15.4$ $\mu $m, which give smaller lattice spacing $\Delta
k_{x}=8.24\times 10^{3}$ m$^{-1}$, together with $\Delta k_{y}=\Delta
k_{z}=4.08\times 10^{5}$ m$^{-1}$, $k_{x}^{(\max )}=1.75\times 10^{7}$ m$%
^{-1}$, and $k_{y}^{(\max )}=k_{z}^{(\max )}=8.16\times 10^{6}$ m$^{-1}$.
Our results on the new lattice reproduce the previous ones, within the
sampling errors of the stochastic simulations. We typically average over $%
2800$ stochastic trajectories, and take $128$ time steps in the simulations
over $25$ $\mu $s collision time. A typical simulation of this size takes
about $100$ hours on $7$ CPUs running in parallel at $3.6$ GHz clock speed.

\section*{References}

\bibliographystyle{prsty}

\bibliography{He_collisionbib}

\end{document}